\documentclass[fleqn,usenatbib]{mnras}

\usepackage{newtxtext,newtxmath}

\usepackage[T1]{fontenc}

\DeclareRobustCommand{\VAN}[3]{#2}
\let\VANthebibliography\thebibliography
\def\thebibliography{\DeclareRobustCommand{\VAN}[3]{##3}\VANthebibliography}

\newcommand\UTUL{H0}
\newcommand\UTL{H014}
\newcommand\URLow{H025}
\newcommand\URHigh{H075}
\newcommand\URVH{H150}
\newcommand\CTUL{F0}
\newcommand\CTL{F014}
\newcommand\CRLow{F025}
\newcommand\CRHigh{F075}
\newcommand\CRVH{F150}
\newcommand\URHighA{H075A}
\newcommand\CRHighA{F075A}
\newcommand\defwidth{0.4}

\usepackage{graphicx}	
\usepackage{amsmath}	
\usepackage{float}
\graphicspath{{./figures/}}

\title[Early dynamics of star clusters]{Early dynamics and violent relaxation of multi-mass rotating star clusters}

\author[A. Livernois et al.]{
Alexander Livernois,$^{1}$\thanks{E-mail: allivern@iu.edu}
Enrico Vesperini,$^{1}$
Maria Tiongco,$^{2}$
Anna Lisa Varri,$^{3,4}$
Emanuele Dalessandro$^5$
\\
$^{1}$Department of Astronomy, Indiana University, Bloomington, IN, 47405, USA\\
$^{2}$University of Colorado, JILA and Department of Astrophysical and Planetary Sciences, 440 UCB, Boulder, CO 80309- USA\\
$^{3}$Institute for Astronomy, University of Edinburgh, Royal Observatory, Blackford Hill, Edinburgh EH9 3HJ, UK\\
$^{4}$School of Mathematics and Maxwell Institute for Mathematical Sciences, University of Edinburgh, Kings Buildings, Edinburgh EH9 3FD, UK\\
$^{5}$ INAF—Astrophysics and Space Science Observatory Bologna, Via Gobetti 93/3 I-40129 Bologna, Italy\\
}

\pubyear{2021}

\begin{document}

\label{firstpage}
\pagerange{\pageref{firstpage}--\pageref{lastpage}}
\maketitle

\begin{abstract}
We present the results of a study aimed at exploring, by means of N-body simulations, the evolution of rotating multi-mass star clusters during the violent relaxation phase, in the presence of a weak external tidal field. We study the implications of the initial rotation and the presence of a mass spectrum for the violent relaxation dynamics and the final properties of the equilibria emerging at the end of this stage. Our simulations show a clear manifestation of the evolution towards spatial mass segregation and evolution towards energy equipartition during and at the end of the violent relaxation phase. We study the final rotational kinematics and show that massive stars tend to rotate more rapidly than low-mass stars around the axis of cluster rotation. Our analysis also reveals that during the violent relaxation phase, massive stars tend to preferentially segregate into orbits with angular momentum aligned with the cluster’s angular momentum, an effect previously found in the context of the long-term evolution of star clusters driven by two-body relaxation.
\end{abstract}

\section{Introduction}

Many recent observational studies have significantly improved our understanding of the dynamical properties of star clusters and provided  key constraints for the study of the dynamical evolution of these systems and the role played by different dynamical processes.
In particular, the last several years have seen a significant leap forward in our understanding of the clusters' internal kinematic properties. Studies based on large samples of radial velocities provided by multi-object spectroscopic surveys (see e.g. \citealt{2014FaMa}, \citealt{2018FeMu}, 
\citealt{2018BaHi}, \citealt{2018KaHu}) and/or high-precision proper motion studies based on Gaia and HST data (see e.g. \citealt{2015Wava}, \citealt{2017BeBi}, \citealt{2018LIBe}, \citealt{2018Biva}, \citealt{2019JiWe}, \citealt{2019Va}, \citealt{2020CoMi}, \citealt{2021CoBe}) have allowed to investigate the velocity dispersion's variation with the cluster-centric distance, have shown that clusters are often characterized by anisotropy in the velocity distribution and/or internal rotation, and provided the first quantitative measures of the degree of evolution towards energy equipartition.

The observational study of young star clusters has also received growing interest in the last several years (see e.g. \citealt{2014mystix}, \citeyear{2015mystixa}, \citeyear{2015mystixb}, \citeyear{2017mystix}, \citealt{2018mystix}),
and thanks to data from the Gaia mission, significant progress has been made in the study of their internal kinematic properties (see e.g. \citealt{2019KuHi}, \citealt{2019GeFe}, \citealt{2020LiHo}, \citealt{2021DaVa}, \citealt{2021SwDO})

On the theoretical side, renewed attention has been paid to the evolution of the clusters' kinematic properties  both during the very early evolutionary phases when the systems  are undergoing so-called violent relaxation and are evolving towards a virial equilibrium state, and during the long-term evolution driven mainly by the effects of two-body relaxation and the host galaxy tidal field.
The role of the dynamical processes acting during the early and long-term evolution of star clusters in determining their present-day properties depends on the clusters' internal structural properties (such as mass and size), their orbit in the Galaxy, the strength of the external tidal field, and their physical and dynamical ages.

Studies of clusters' formation and very early evolution have shown that clusters can emerge from these evolutionary phases with dynamical properties characterized by mass segregation, internal rotation, and radial anisotropy in the velocity distribution; the detailed role of different dynamical processes acting during these phases and the variety of different dynamical paths resulting in these dynamical properties are still matter of intense investigation (see e.g. \citealt{2004GoWh}, \citealt{2007McVe}, \citealt{2009AlGo}, \citeyear{2010AlGo}, \citealt{2009MoBo}, \citealt{2012FuSa}, \citealt{2016FuPo}, \citealt{2014VeVa}, \citealt{2016PaGo}, \citealt{2017DoFe}, \citealt{MapMi},
\citealt{2014BaKr}, \citeyear{2017BaKr}, \citeyear{2018BaKr}, \citealt{2018SiRi}, \citealt{2020DaPa}, \citealt{2020BaMa}, \citealt{2021BaTo}).

For old star clusters such as massive globular clusters, the dynamical properties are either entirely due to  the effects of internal two-body relaxation and the interplay between relaxation and the external tidal field of the Galaxy or the result of how relaxation and external tidal field have altered the dynamical properties imprinted at the time of their formation and during the early dynamical evolution.
Several studies, for example, have shown that the cluster's initial internal rotation is gradually erased by the effects of two-body relaxation and the loss of angular momentum carried away by stars escaping  during a cluster long-term evolution (see e.g. \citealt{1999EiSp}, \citealt{2007ErGl}, \citealt{2013HoKi}, \citealt{2017TiVe}, \citealt{2018KaHu}). The strength of the present-day rotation thus represents a lower-limit on the initial cluster's rotation. Part of this rotation may be due to a partial spin-orbit synchronization resulting from the coupling between the cluster's internal kinematics and the effects of the external tidal field (\citealt{frac_ret}, \citeyear{2019TiVe}, \citealt{2018LaFe}, \citealt{2021DaRa}, \citealt{2021WaOl}).
As for the anisotropy of the velocity distribution in both the early and the long-term dynamical history, the cluster's initial structural properties, its dynamical age, and the extent of the mass loss suffered by a cluster during its evolution can play key roles in determining the evolution of the velocity anisotropy and its present-day strength (see e.g. \citealt{2014VeVa}, \citealt{2016aTiVe}).

In this paper, we present the results of a suite of N-body simulations aimed at exploring the dynamics of rotating clusters during the violent relaxation phase, in the presence of a weak external tidal field. This study is focused, in particular, on the role and the dynamical implications of the initial rotation for the structural and kinematic properties of the final equilibria emerging at the end of the violent relaxation phase and on how these properties depend on the star masses. Understanding the clusters' early evolutionary stages and characterizing the dynamical properties imprinted during those phases are critical steps to be able to build a theoretical framework for the interpretation of the observed  properties of young clusters and to reconstruct the evolutionary history of dynamically old clusters.

The outline of the paper is the following: in Section \ref{Methods} we present our methods and initial conditions, Section \ref{Results} is devoted to the presentation of the results, and in Section \ref{Conclusions} we summarize our conclusions.

\section{Methods and Initial Conditions}
\label{Methods}

\begin{table}

\centering
\begin{tabular}{|l|l|l|}
\hline
ID.  & Initial Spatial Distribution & $Q_{\rm rot}$ \\ \hline
\UTUL\ & Homogeneous & 0         \\ 
\UTL\   & Homogeneous & 0.014    \\ 
\URLow\ & Homogeneous & 0.025 \\ 
\URHigh\ & Homogeneous & 0.075   \\
\URVH& Homogeneous   & 0.15  \\
\CTUL\ & Fractal     & 0     \\
\CTL\   & Fractal     & 0.014\\
\CRLow\ & Fractal     & 0.025\\
\CRHigh\ & Fractal    & 0.075  \\
\CRVH\ & Fractal      & 0.15 \\ \hline
\end{tabular}

\caption{Summary of initial conditions for the N-body simulations presented in this paper. The rotational virial ratio, $Q_{\rm rot}$, is the ratio of the total initial rotational kinetic energy from the added solid-body, bulk rotation to the potential energy (see equation \ref{eq:virial}). For all the systems, the initial virial ratio associated with random motion is equal to 0.001. \UTL\  and \CTL\  are initially tidally locked (i.e. their initial internal rotation is synchronized with the orbital rotation of the cluster around the host galaxy) .}
\label{SimTable}
\end{table}

\begin{figure*}
    \centering
    \includegraphics[width=0.9\textwidth]{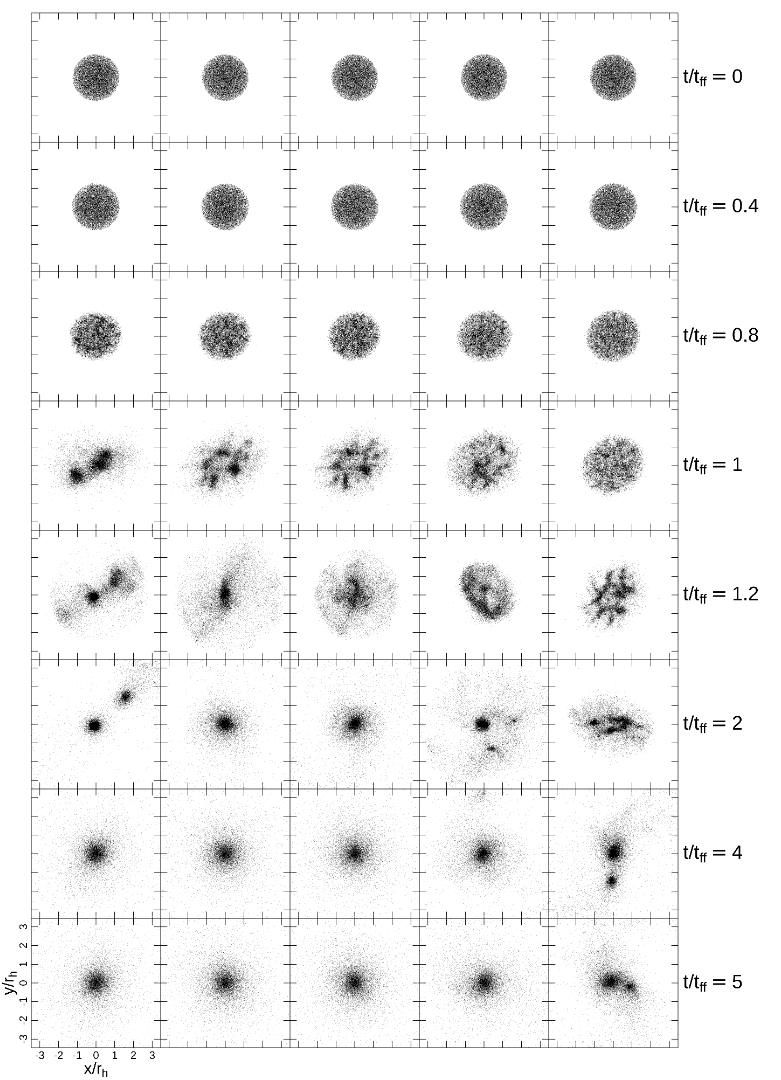}
    \caption{Time evolution of the spatial distributions in the $x$-$y$ plane for each initially homogeneous cluster (from left to right: \UTUL, \UTL, \URLow, \URHigh, \URVH). The coordinates $x$ and $y$ are scaled to the 3-D half-mass radius, $r_{\rm h}$, evaluated for the system at the time of the snapshot. The sample of times chosen provide a representative illustration of the clusters' structural evolution.}
    \label{fig:uni_struct_xy}
\end{figure*}
\begin{figure*}
    \centering
    \includegraphics[width=0.9\textwidth]{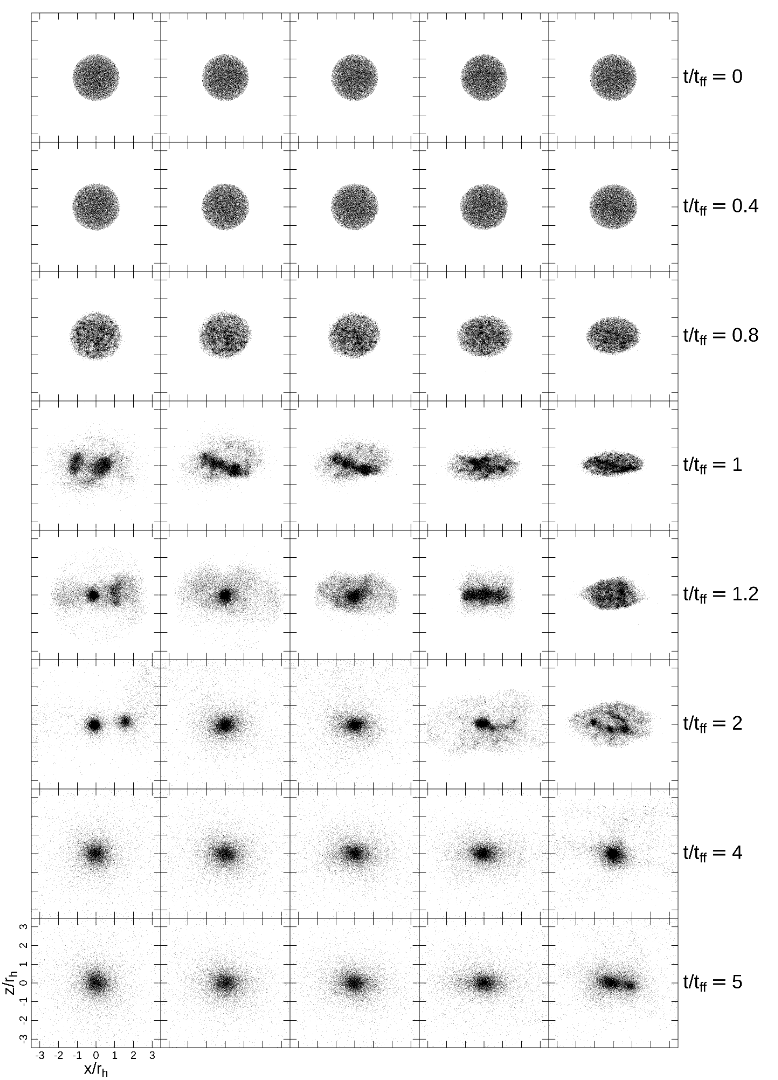}
      \caption{Time evolution of the spatial distributions in the $x$-$z$ plane for each initially homogeneous cluster (from left to right: \UTUL, \UTL, \URLow, \URHigh, \URVH). The coordinates $x$ and $z$ are scaled to the 3-D half-mass radius, $r_{\rm h}$, evaluated for the system at the time of the snapshot. These show that the clusters with highest rotation have more clumps that survive the maximum-contraction stages.}
    \label{fig:uni_struct_xz}
\end{figure*}
\begin{figure*}
    \centering
    \includegraphics[width=0.9\textwidth]{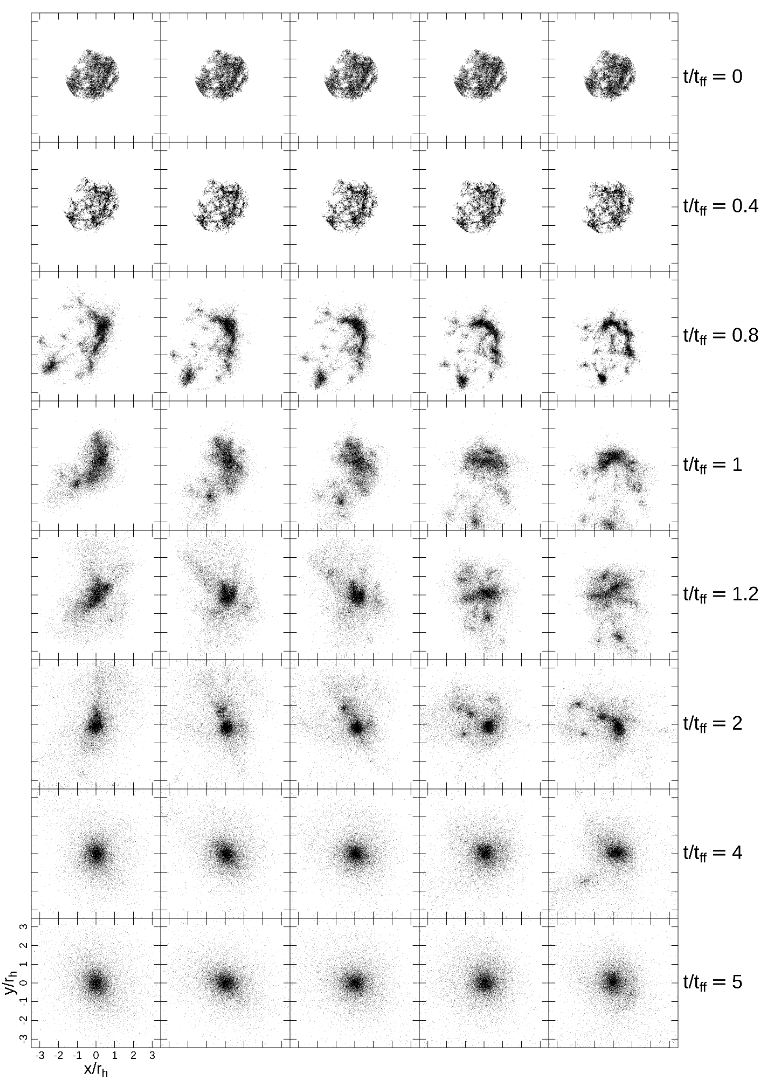}
    \caption{Time evolution of the spatial distributions in the $x$-$y$ plane for all the models with an initial fractal spatial distribution (from left to right: \CTUL, \CTL, \CRLow, \CRHigh, \CRVH). The coordinates $x$ and $y$ are scaled to the 3-D half-mass radius, $r_{\rm h}$, evaluated for the system at the time of the snapshot. These show that the initially fractal clusters form larger scale clumps than the initially homogeneous clusters before maximum contraction.}
    \label{fig:clumpy_struct_xy}
\end{figure*}
\begin{figure*}
    \centering
    \includegraphics[width=0.9\textwidth]{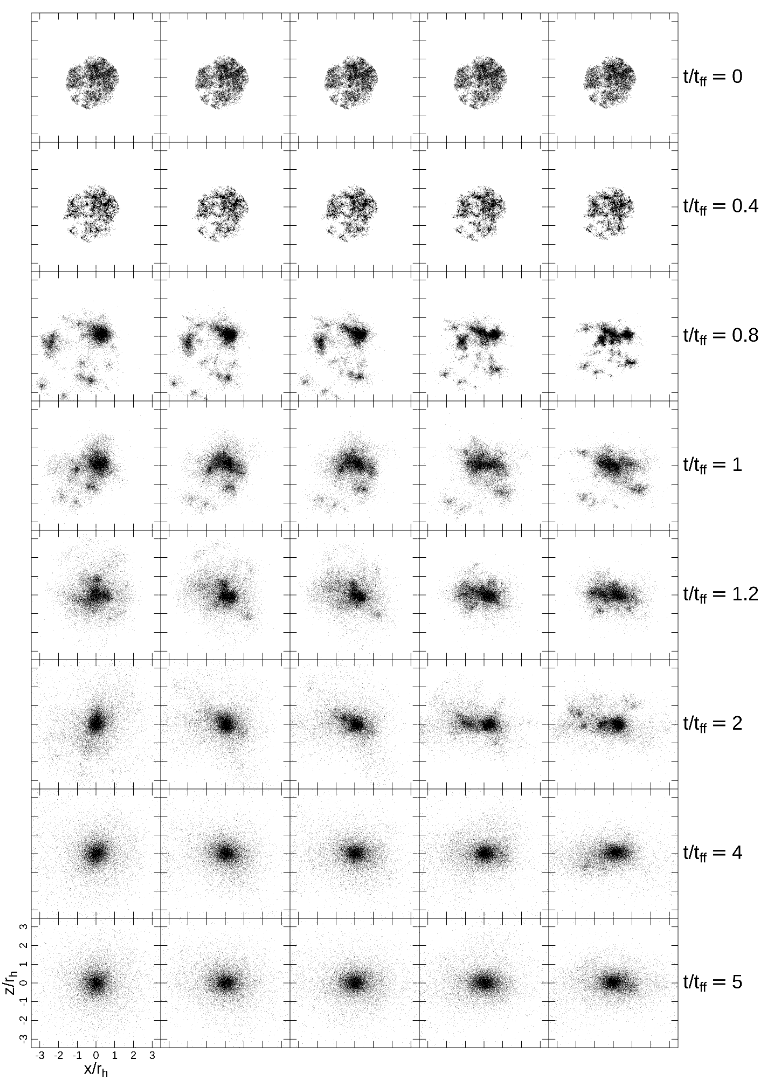}
    \caption{Time evolution of the spatial distributions in the $x$-$z$ plane for each initially fractal cluster (from left to right: \CTUL, \CTL, \CRLow, \CRHigh, \CRVH). The coordinates $x$ and $z$ are scaled to the 3-D half-mass radius, $r_{\rm h}$, evaluated for the system at the time of the snapshot. The sample of times are chosen as representative of the structural evolution. These show that the clusters with highest rotation have more clumps that survive the maximum-contraction stages.}
    \label{fig:clumpy_struct_xz}
\end{figure*}
\begin{figure*}
\centering

\includegraphics[width=\defwidth\textwidth]{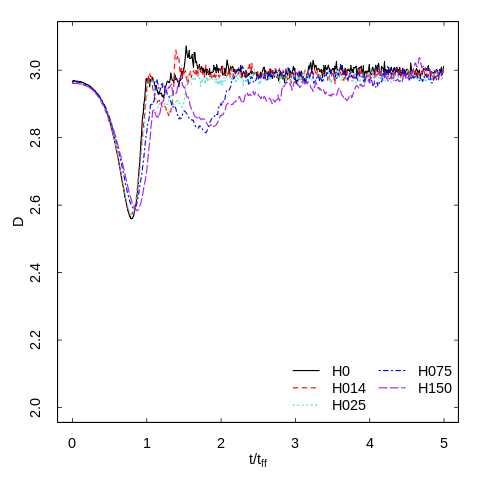}
\includegraphics[width=\defwidth\textwidth]{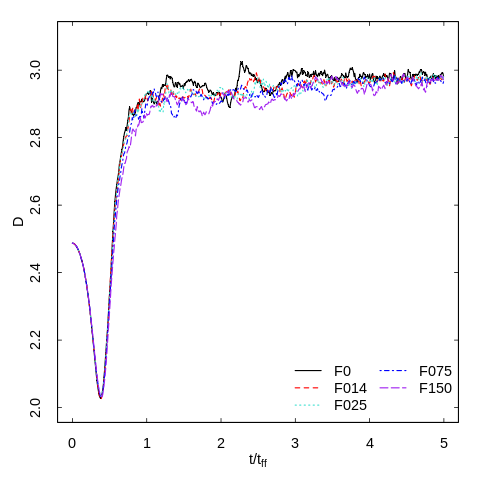}

\caption{Time evolution of the fractal dimension, $D$ (see equation \ref{eq:D}), for the initially homogeneous clusters (left panel) and the initially fractal clusters (right panel). This displays the level of fractal structure present in all clusters, and that the clusters with higher initial rotation maintain their substructure for longer before evolving into a monolithic structure.}
\label{fig:fractal_comb}
\end{figure*}

\begin{figure*}
    \centering
    \includegraphics[width=0.4\textwidth]{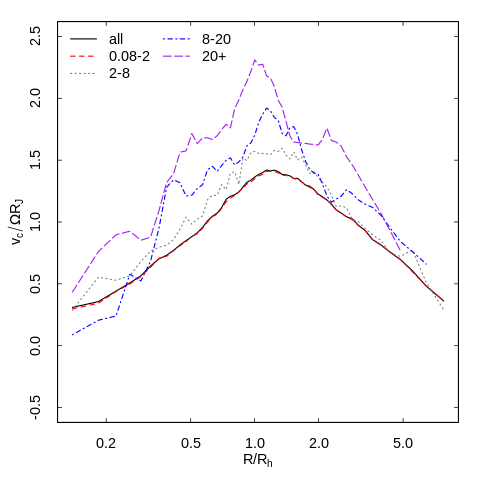}
    \includegraphics[width=0.4\textwidth]{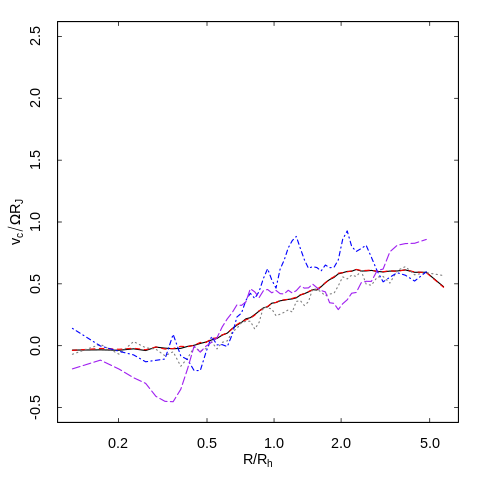}
    \includegraphics[width=0.4\textwidth]{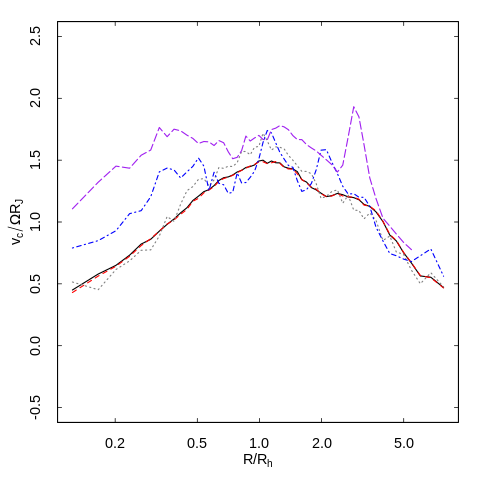}
    \includegraphics[width=0.4\textwidth]{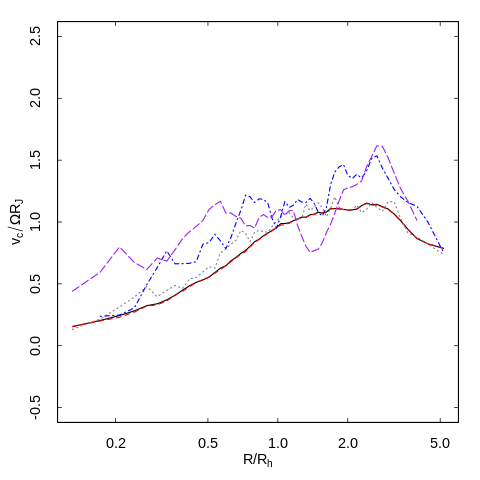}
    \includegraphics[width=0.4\textwidth]{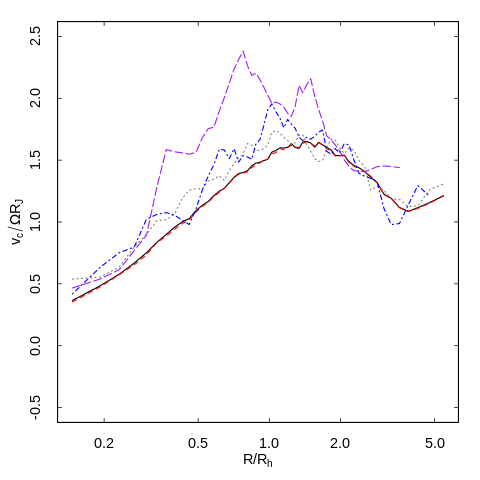}
    \includegraphics[width=0.4\textwidth]{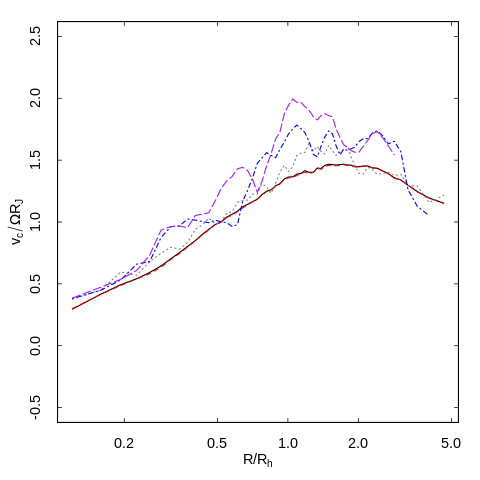}
    \caption{Radial profile of the rotation velocity, $v_{\rm c}$, for each mass range in the legend (in units of $M_{\sun}$), for the \URLow, \URHigh, and \URVH\ (top to bottom, left panel) clusters, and \CRLow, \CRHigh, \CRVH\ (top to bottom, right panel) clusters from $4.5 t_{\rm ff}<t<5 t_{\rm ff}$ (except for \URVH\ and \CRVH\, which are evaluated over $5.5 t_{\rm ff}<t<6 t_{\rm ff}$, since they reach equilibrium later). $R$ and $R_{\rm h}$ are, respectively, the cylindrical radius and cylindrical half-mass radius; $\Omega$ is the angular velocity of the cluster's orbital motion around the host galaxy and $R_{\rm J}$ is the cluster's Jacobi radius. ($\Omega R_{\rm J}$ represents the rotational velocity at the Jacobi radius for a system with internal rotation synchronized with the orbital rotation around the host galaxy).}
    \label{fig:Vcurve_low}
\end{figure*}
\begin{figure*}
    \centering
    \includegraphics[width=\textwidth]{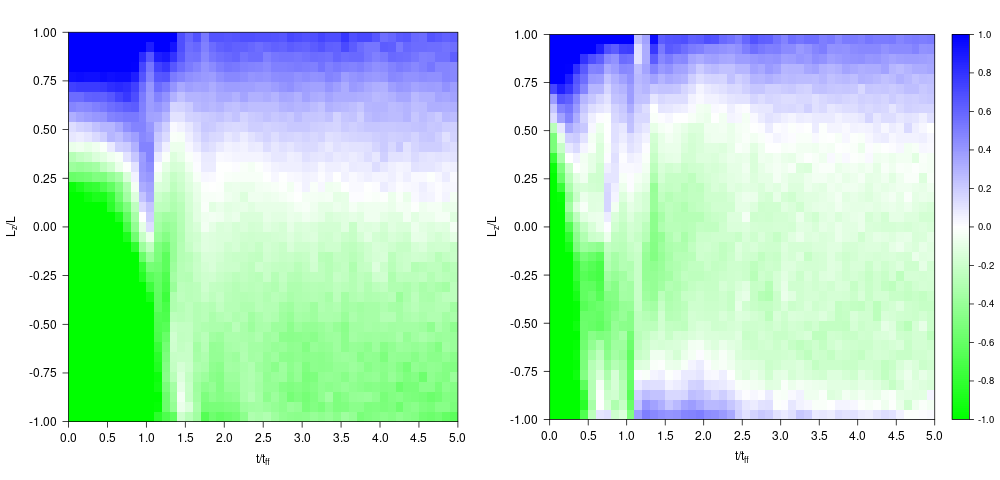}
    \caption{Time evolution of the distribution of the orbital alignment, $L_{\rm z}/L$, in the inner 50\% in radius of stars in \URHigh\ (left panel) and \CRHigh\ (right panel). The colour is defined by $\ln(N(L_{\rm z}/L,t)/N(t))$. These show clear orbital alignment with the system's initial rotation for the initially homogeneous cluster (left panel), and a bi-modal distribution of the orbital alignment in the initially fractal cluster (right panel), indicating the presence of a counter-rotating subsystem.}
    \label{fig:LzLEvo_all}
\end{figure*}

\begin{figure*}
    \centering
    \includegraphics[width=0.8\textwidth]{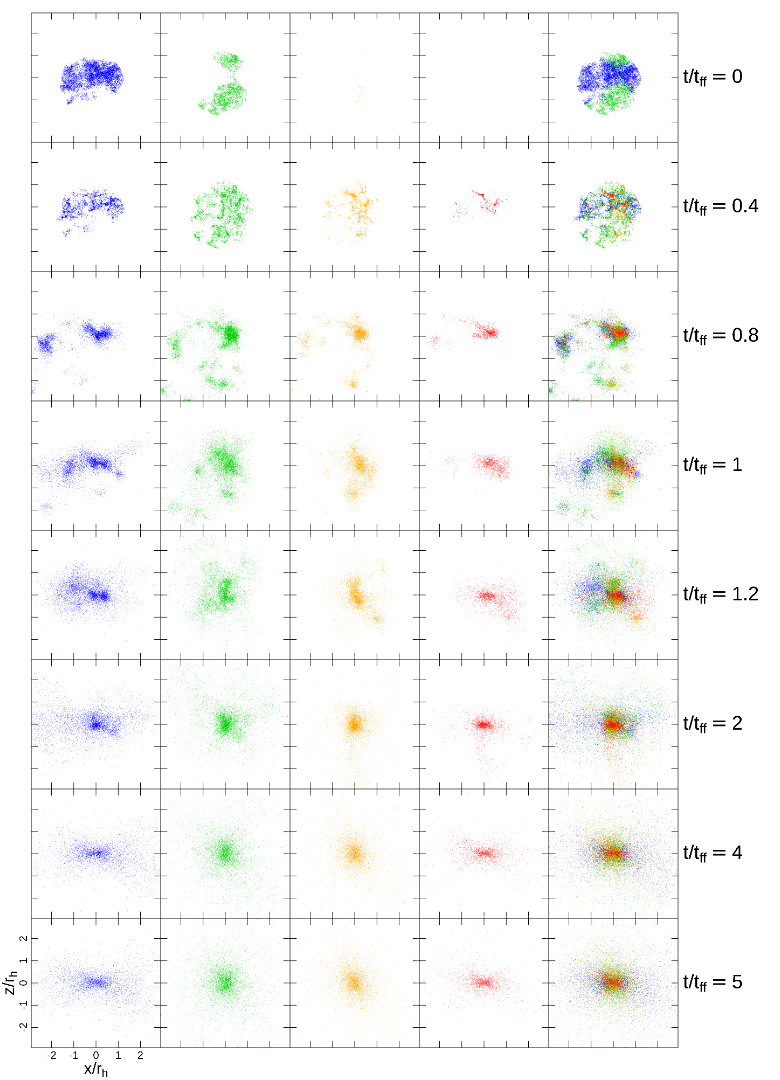}
    \caption {Different dynamical populations of the \CRLow\ cluster in the $x$-$z$ plane. The columns are (from left to right): $L_{\rm z}/L>0.75$, $0<L_{\rm z}/L<0.75$, $-0.75<L_{\rm z}/L<0$, $L_{\rm z}/L<-0.75$, and $-1\le L_{\rm z}/L\le 1$, with each subgroup keeping their original colour. The coordinates $x$ and $z$ are scaled to the 3-D half-mass radius, $r_{\rm h}$, evaluated for the system at the time of the snapshot. This shows the formation of a counter-rotating inner disc, as well as the dynamics of the larger clumps during the violent relaxation phase. At $t=5t_{\rm ff}$, the percentage of stars in each angular alignment bin (from left to right) is approximately as follows: 24\%, 35\%, 29\%, 12\%, 100\%.}
    \label{fig:ret_framexz_CRLow}
\end{figure*}

\begin{figure*}
    \centering
    \includegraphics[width=\defwidth\textwidth]{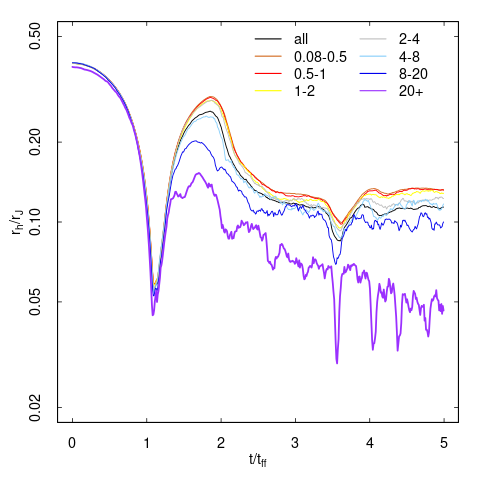}
    \includegraphics[width=\defwidth\textwidth]{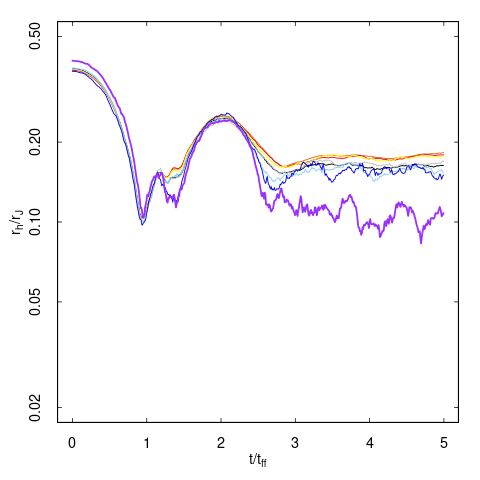}
    \caption{Time evolution of the 3-D half-mass radius, $r_{\rm h}$, normalized by the Jacobi radius, $r_{\rm J}$, for each mass range in the legend (in units of $M_{\sun}$), for \URHigh\ (left panel) and \CRHigh\ (right panel). The trend between $r_{\rm h}$ and stellar mass clearly shows evidence of mass segregation after around 1 $t_{\rm ff}$.}
    \label{fig:HMRadii}
\end{figure*}

\begin{figure*}
\centering
\includegraphics[width=\defwidth\textwidth]{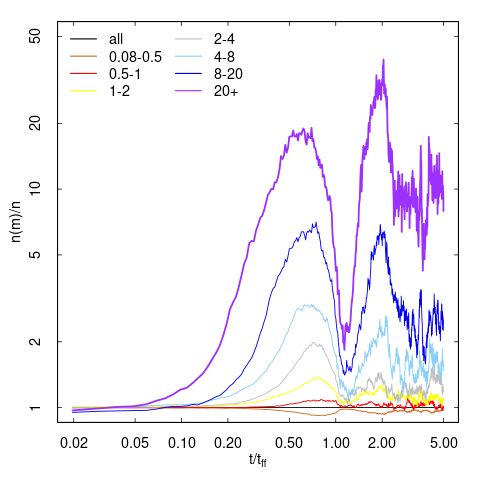}
\includegraphics[width=\defwidth\textwidth]{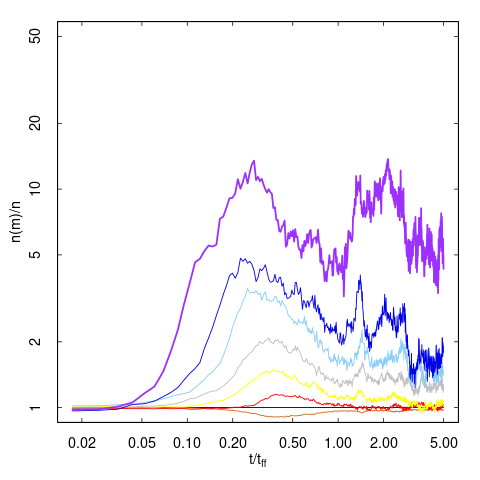}
\caption{Time evolution of the median local number density per mass group, $n(m)$, for each mass range in the legend (in units of $M_{\sun}$), normalized by the median local density of the entire cluster, $n$, for \URHigh\ (left panel) and \CRHigh\ (right panel). The trend between $n(m)$ and particle mass clearly shows evidence of mass segregation well before maximum contraction at $\approx1t_{\rm ff}$.}
\label{fig:MSeg}
\end{figure*}

\begin{figure*}
\centering
\includegraphics[width=\defwidth\textwidth]{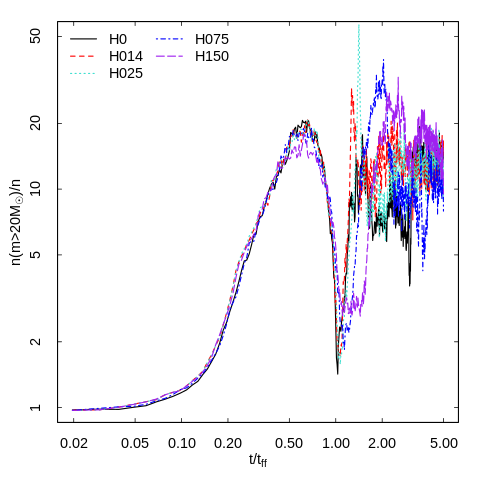}
\includegraphics[width=\defwidth\textwidth]{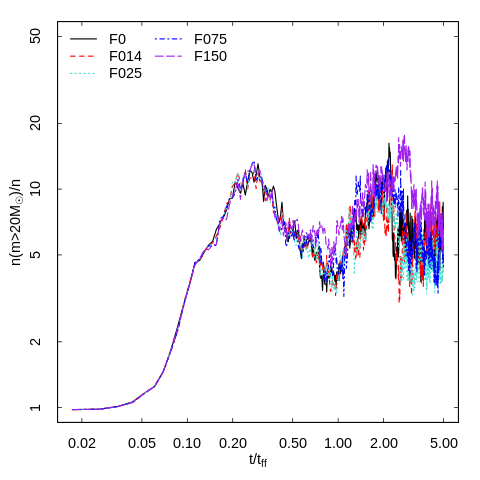}

\caption{Time evolution of the median local number density for massive stars, $n(m>20M_{\odot})$, normalized by the median local number density for the entire cluster, $n$, for the initially homogeneous clusters (left panel) and the initially fractal clusters (right panel). This shows that the early mass segregation is consistent between all clusters, and more significant in the initially homogeneous clusters.}
\label{fig:Mseg_comb}
\end{figure*}

\begin{figure}
    \centering
    \includegraphics[width=\defwidth\textwidth]{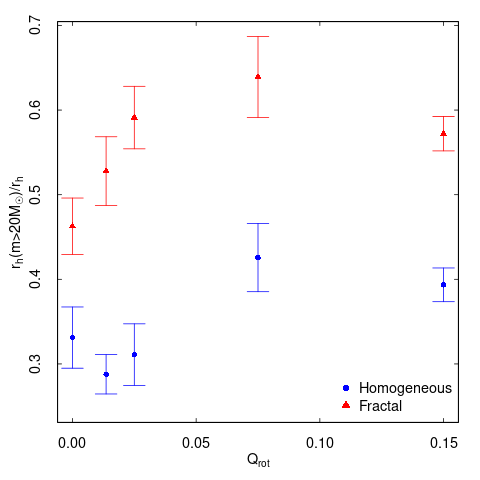}
    \caption{Half-mass radius, $r_{\rm h}$, for stars with $m>20M_{\odot}$, normalized by the half-mass radius of the entire cluster and averaged over $4.5 t_{\rm ff}<t<5 t_{\rm ff}$ as a function of initial rotation, for each cluster. The error bars are the standard deviation of the averaged snapshots. This shows that once the cluster reaches equilibrium, the initially homogeneous clusters are characterized by stronger mass segregation than the initially fractal systems, and also that rotation has a significant effect on the final mass segregation.}
    \label{fig:HMR_late}
\end{figure}

\begin{figure*}
    \centering
    \includegraphics[width=\defwidth\textwidth]{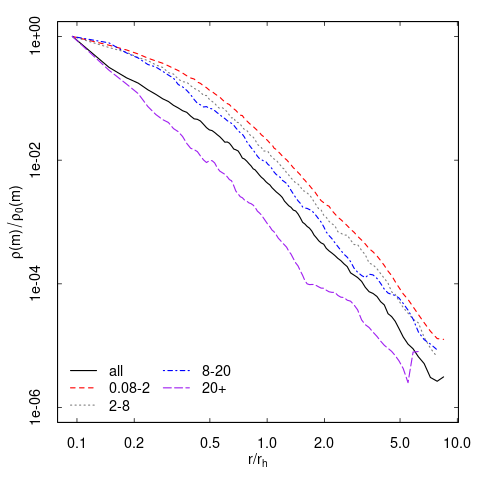}
    \includegraphics[width=\defwidth\textwidth]{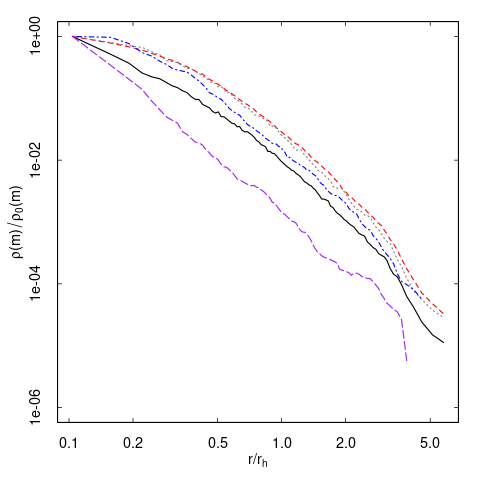}
    \caption{Median radial profile of the mass density, $\rho$, for each mass range in the legend (in units of $M_{\sun}$), normalized by the mass density of each mass range in the innermost radial bin, $\rho_{0}$, for the \URHigh\ (left panel) and \CRHigh\ (right panel) clusters from $4.5 t_{\rm ff}<t<5 t_{\rm ff}$. In the inner regions, the high-mass stars display a more cusp-like profile as opposed to the more core-like profile of the less massive stars.}
    \label{fig:Dens_mass}
\end{figure*}

\begin{figure*}
    \centering
    \includegraphics[width=0.8\textwidth]{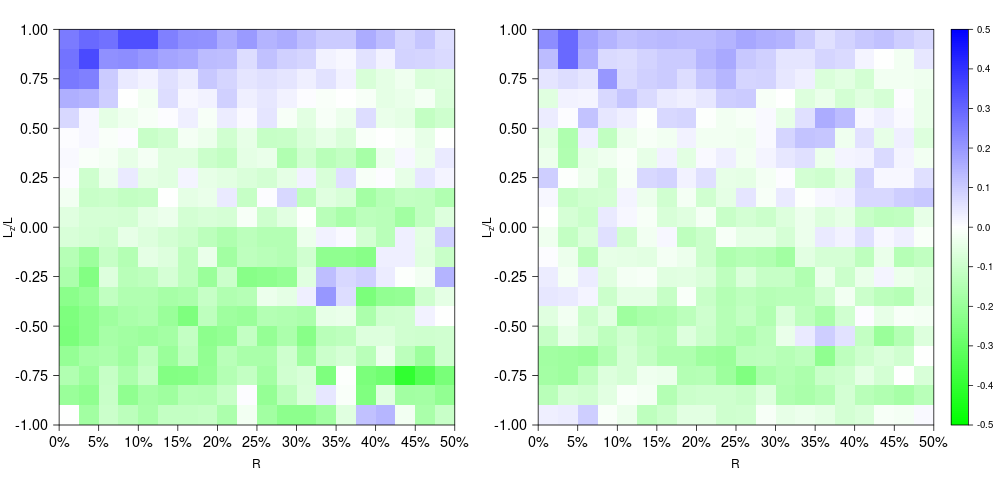}
    \caption{Enhancement of average mass, $\epsilon$ (defined in equation \ref{eq:mass_en}), of different Lagrangian cylindrical radii and $L_{\rm z}/L$ bins for \URHigh\ (left panel) and \CRHigh\ (right panel). The colour is determined by $\ln(\epsilon)$. The snapshots with $4 t_{\rm ff}<t<5 t_{\rm ff}$ are combined in this analysis to improve statistics. The blue region around $L_{\rm z}/L\approx1$ for both clusters displays how the massive stars are preferentially on orbits rotating about the cluster rotation axis as compared to less massive stars.}
    \label{fig:AvgMassEn}
\end{figure*}

\begin{figure*}
\centering
\includegraphics[width=\defwidth\textwidth]{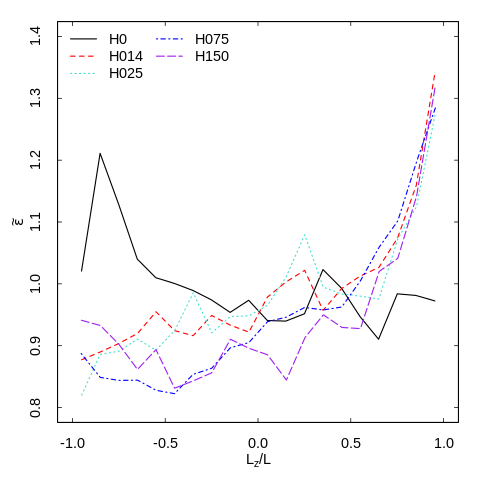}
\includegraphics[width=\defwidth\textwidth]{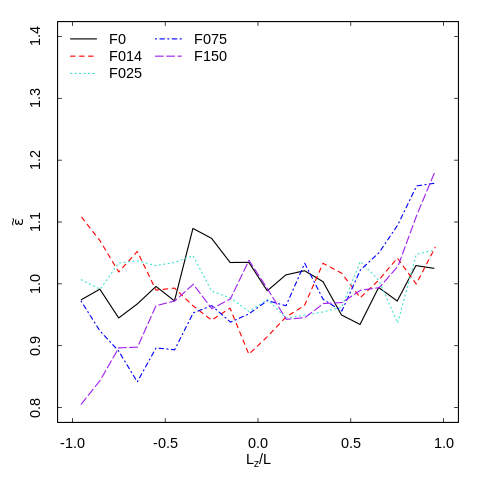}
\caption{Enhancement of average mass for the innermost 25\% of stars, $\tilde\epsilon$ (defined in equation \ref{eq:mass_en_eff}), as a function of orbital orientation, $L_{\rm z}/L$, at equilibrium for the initially homogeneous clusters (left panel) and the initially fractal clusters (right panel). For all clusters except for \URVH\ and \CRVH\, $\tilde\epsilon$ is calculated between 4-5$t_{\rm ff}$, and for the \URVH\ and \CRVH\ clusters, 5.5-6.5 $t_{\rm ff}$, due to those clusters reaching equilibrium later.}
\label{fig:AvgMassEn_table}
\end{figure*}

\begin{figure*}
    \centering
    \includegraphics[width=\defwidth\textwidth]{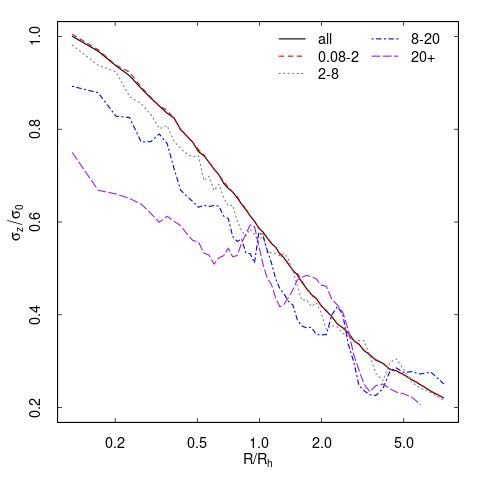}
    \includegraphics[width=\defwidth\textwidth]{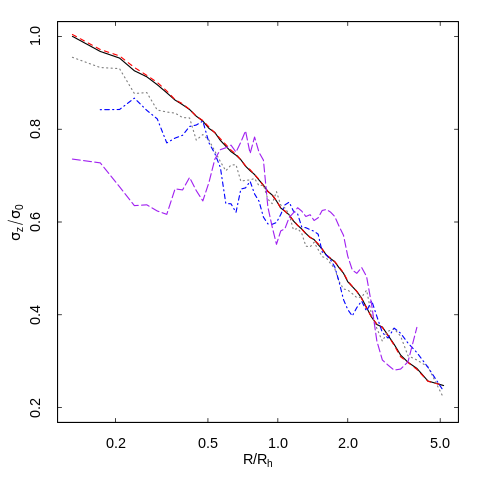}
    \caption{Radial profile of the velocity dispersion along the $z$-axis, $\sigma_{\rm z}$, for each mass range in the legend (in units of $M_{\sun}$), normalized by the central velocity dispersion of the entire cluster along the same axis, $\sigma_{0}$, for the \URHigh\ (left panel) and \CRHigh\ (right panel) clusters from $4.5 t_{\rm ff}<t<5 t_{\rm ff}$. $R_{\rm h}$ is the cylindrical half mass radius of the system. The clear trend between the inner $\sigma_{\rm z}$ and mass is evidence of early evolution towards energy equipartition in these systems.}
    \label{fig:DispvsR}
\end{figure*}

\begin{figure*}
    \centering
    \includegraphics[width=\defwidth\textwidth]{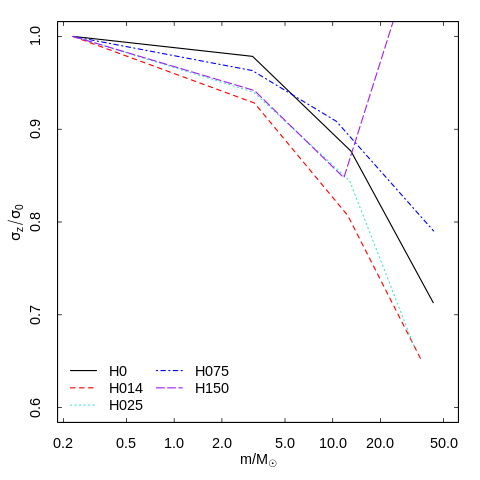}
    \includegraphics[width=\defwidth\textwidth]{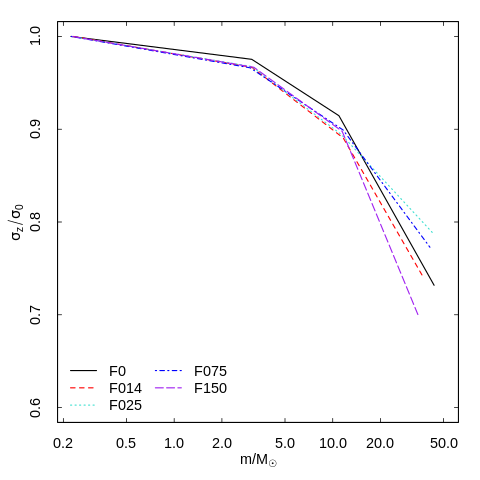}
    \caption{Velocity dispersion along the $z$-axis, $\sigma_{\rm z}$, normalized by the central velocity dispersion along the same axis, $\sigma_{z,0}$, as a function of mass for the inner 10\% in radius of stars for $4.5 t_{\rm ff}<t<5 t_{\rm ff}$ (except for \URVH\ and \CRVH\, which are evaluated over $5.5 t_{\rm ff}<t<6 t_{\rm ff}$, since they reach equilibrium later) for the initially homogeneous clusters (left panel) and the initially fractal clusters (right panel). This shows the extent of evolution towards energy equipartition reached in each cluster by the time they reach equilibrium.}
    \label{fig:EEqui}
\end{figure*}

\begin{figure}
    \centering
    \includegraphics[width=\defwidth\textwidth]{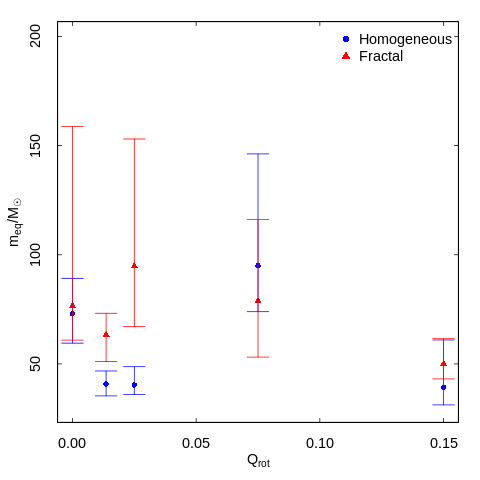}
    \caption{The equipartition mass, $m_{\rm eq}$ (defined in equation \ref{eq:meq_sig}), for each cluster for the inner 10\% in radius of stars as a function of $Q_{\rm rot}$ over $4.5 t_{\rm ff}<t<5 t_{\rm ff}$ (except for \URVH\ and \CRVH\, which are evaluated over $5.5 t_{\rm ff}<t<6 t_{\rm ff}$, since they reach equilibrium later). Due to the anomalous behavior of the 20$M_{\sun}+$ mass bin in the \URVH\  curve in Fig. \ref{fig:EEqui}, we omit that mass bin for our calculation of $m_{\rm eq}$.}
    \label{fig:EEqui_mass}
\end{figure}

\begin{figure*}
    \centering
    \includegraphics[width=\defwidth\textwidth]{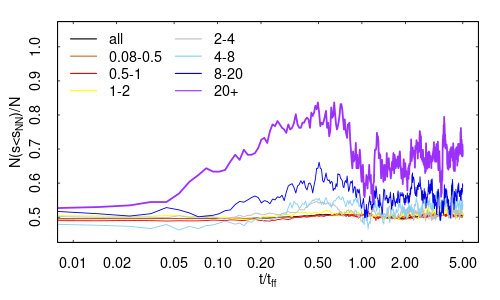}
    \includegraphics[width=\defwidth\textwidth]{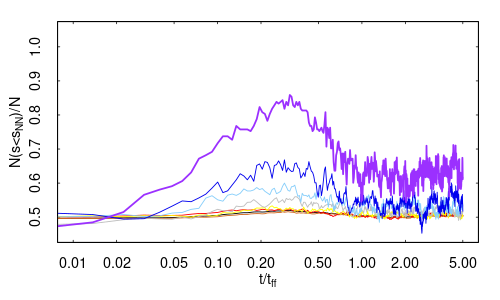}
    \includegraphics[width=\defwidth\textwidth]{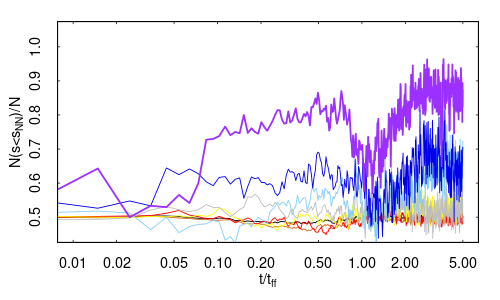}
    \includegraphics[width=\defwidth\textwidth]{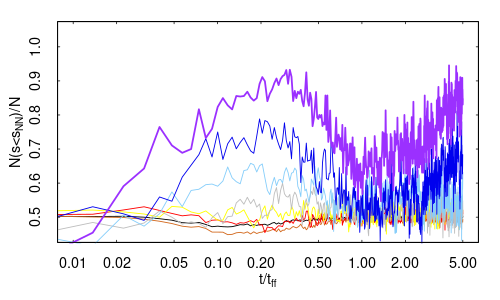}
    \caption{Time evolution of the fraction of stars that have a speed lower than the median speed of their nearest 20 neighbours for stars, $N(s<s_{\rm NN})/N$, with all local density values (top row), and for stars with the highest 10\% values in local density (bottom row) for each mass range in the legend (in units of $M_{\sun}$) in \URHigh\ (left panels) and \CRHigh\ (right panels). This shows how early in the violent relaxation process that the evolution towards energy equipartition starts.}
    \label{fig:EEqui_localall}
\end{figure*}

\begin{figure*}
    \centering
    \includegraphics[width=\defwidth\textwidth]{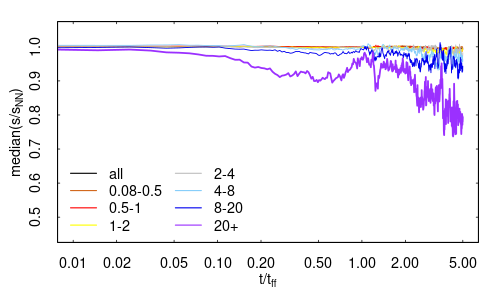}
    \includegraphics[width=\defwidth\textwidth]{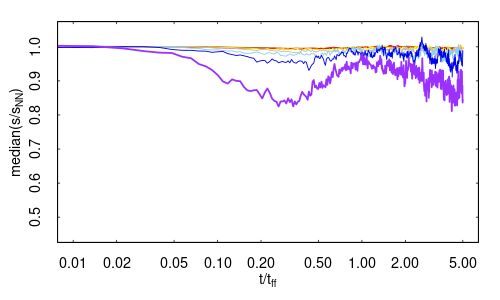}
    \includegraphics[width=\defwidth\textwidth]{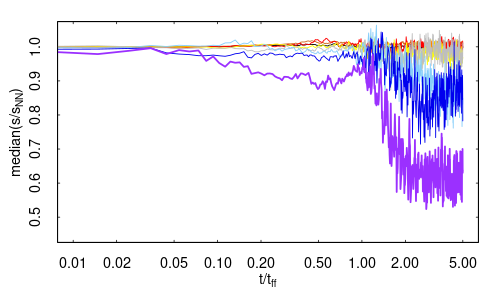}
    \includegraphics[width=\defwidth\textwidth]{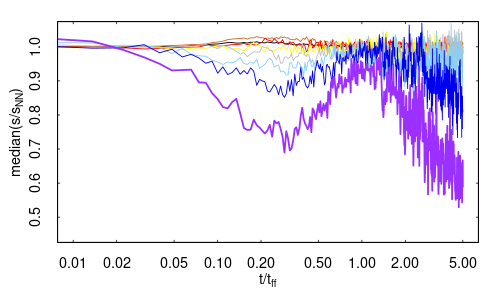}
    \caption{Time evolution of the median local relative speed, $s/s_{\rm NN}$ (where $s$ is the speed of each star and $s_{\rm NN}$ is the median speed of each star's 20 nearest-neighbours), for stars with all local density values (top row), and for stars with the highest 10\% values in local density (bottom row) for each mass range in the legend (in units of $M_{\sun}$) in \URHigh\ (left panels) and \CRHigh\ (right panels).}
    \label{fig:EEqui_div_localall}
\end{figure*}
 
This study is based on a set of ten N-body simulations run using the GPU-accelerated version of the \textsc{NBODY6} code (\citealt{nbody6_2003}, \citealt{GPU_nbody6}) on the Indiana University's Big Red II and Carbonate supercomputers. All the  simulations start with $N = 10^5$ stars with masses distributed according to a \cite{KIMF} initial mass function between 0.08 and 100 solar masses.

Our systems start with no primordial binaries and in these simulations we do not include the effects of stellar evolution. Both of these ingredients will be considered in future studies.

The clusters we have studied are initially characterized by either a homogeneous density profile or a clumpy, fractal spatial distribution with fractal dimension $D\sim 2.5$. The initial fractal distribution was created using the \textsc{McLuster} code \citep{mcluster}.

The fractal dimension $D$ is also used throughout our analysis to provide a quantitative measure of clusters' substructure, where $D\approx3$ is indicative of a monolithic distribution, and a smaller $D$ is associated with hierarchical structure; we calculate $D$ (see  e.g. \citealt{fractalknn}) as the best fit to:

\begin{equation}
    median(d)^2\propto k^{2/D}
    \label{eq:D}
\end{equation}
where $d$ is the distance to the $k$th nearest-neighbour for all stars in the cluster and $k$ ranges from 10 to 100 (all $k$th nearest neighbor analyses are calculated using the \textsc{R} \textsc{spatstat} library \citealt{spatstatBook}).

We include the effects of the host galaxy tidal field, modeled as a point-mass, and we assume the clusters to be on  circular orbits; all the clusters have an initial maximum radius, $r_{\rm tot}$ equal to half of the Jacobi radius $r_{\rm J}$. Stars moving beyond a radius equal to twice the Jacobi radius are removed from the simulations. For our analysis, we study the clusters in an inertial frame of reference. We have chosen to consider a single set-up for the tidal environment because we wanted to focus our analysis on the effect of the interplay between the role of the mass spectrum and the initial angular momentum. We will further explore the evolution in tidal fields with different strengths in a future study.

All the systems investigated have the initial ratio of the total kinetic energy from random motion to the potential energy, $Q_{\rm rand}$, is equal to about $0.001$. Different degrees of internal solid-body rotation around an axis perpendicular to the cluster’s orbital plane are then added to the systems; we quantify the initial rotation by means of the ratio of the total kinetic energy from the added rotation to the total potential energy (denoted by U):

\begin{equation}
    Q_{\rm rot}=
    -\frac{\omega_{z,i}^2 \sum\limits_{n=1}^N m_{n}R_n^2}{2U}
    \label{eq:virial}
\end{equation}

\noindent where $\omega_{z,i}$ is the initial solid-body angular velocity added to the system, $m_n$ is the mass of the nth star, and $R_{\rm n}$ is the cylindrical radius of the $n$th star's location.
The initial values of $Q_{\rm rot}$ explored in our survey are summarized in Table \ref{SimTable} along with the ID we will use in the rest of the paper to refer to each model. In order to quantify the stochastic variations in our results, in addition to the models listed in Table \ref{SimTable}, we have also run two additional simulations starting with different random realizations of the H075 and F075 models. The results of these simulations are presented in Appendix \ref{Appendix}.

The simulations follow the violent relaxation of the clusters until they reach a virial equilibrium state and  are all stopped after about 5 $t_{\rm ff}$, where $t_{\rm ff}$ is the free-fall time defined as:
\begin{equation}
    t_{\rm ff}= \sqrt{\frac{3\pi}{32G\rho_i}}
    =\sqrt{\frac{\pi^2r_{\rm tot,i}^3 }{8GM_{\rm tot,i}}}
\end{equation}

\noindent where $r_{\rm tot,i}$ and $M_{\rm tot,i}$ are the initial total radius and total mass of the cluster, respectively, $\rho_i$ is the initial average mass density of the system, and $G$ is the gravitational constant. The \URVH\ and \CRVH\ clusters do not reach equilibrium until around 5.5 $t_{\rm ff}$, so these models are run until 6.5 $t_{\rm ff}$.
We use $t_{\rm ff}$ for a general characterization of the various phases of the violent relaxation but, as we point out in  the presentation of several results later in the paper, the hierarchical nature of the systems we explore implies that some of the dynamical processes affecting the cluster evolution act on the local timescales of the substructures present during the cluster early evolution.

\section{Results}
\label{Results}
\subsection{Structural properties}
We start the analysis of our results by focusing our attention on the evolution and final properties of the clusters' structure.

Figs. \ref{fig:uni_struct_xy}-\ref{fig:clumpy_struct_xz} show the structural evolution of clusters with different initial density profiles and degrees of rotation during the various phases of violent relaxation including the initial collapse, re-bounce, and the eventual reaching of equilibrium (notice that in each panel the spatial coordinates are normalized to the value of the half-mass radius calculated at the time shown in the panel and that Figs. \ref{fig:uni_struct_xy}-\ref{fig:clumpy_struct_xz} provide a comoving view of the clusters' evolution).

Figs. \ref{fig:uni_struct_xy} and \ref{fig:uni_struct_xz} show the evolution of initially homogeneous systems.
During the early evolutionary phases $(0<t\lesssim t_{\rm ff})$, systems starting with a homogeneous density profile are characterized by the rapid growth of initial density Poisson fluctuations and the formation of several distinct stellar clumps (see also \citealt{1988AaLi} for some of the first simulations showing evidence of this early clump growth). 
Comparing systems with different initial rotations shows that these clumps are less prominent in models with high rotation and clearly illustrate the role played by rotation in these early evolutionary phases.

Most of the stellar clumps formed during the early collapse merge during the maximum contraction phase $(t\sim t_{\rm ff})$, but some may survive; the post-maximum-contraction phase is characterized by the presence of two or three large stellar subsystems orbiting around each other and merging to form a single star cluster only later in the evolution. In the post-maximum-contraction phase, the more tangentially-biased velocity distribution of systems with initial rotation can delay the merging of the stellar clumps that survive the maximum contraction phase and extend the time required for a system to reach a final monolithic structure.
The role of rotation is also evident in the increasing flattening along the direction of the rotation axis (the $z$-axis) for clusters with higher initial rotation (see e.g. Fig. \ref{fig:uni_struct_xz} at $t/t_{\rm ff}=1$).

Figs. \ref{fig:clumpy_struct_xy} and \ref{fig:clumpy_struct_xz}  show the structural evolution of systems with an initially fractal spatial distribution. The role of rotation is similar to that found for initially homogeneous systems. In this case, however, rotation has an additional effect on the early evolution of the stellar clumps present in the initial spatial distribution.  In systems with no or low initial rotation, these subsystems tend to interact with each other, merge, and form larger substructures quickly during the early phases of the system's collapse; these early mergers are increasingly suppressed for larger initial rotations. The result is that initially rotating systems are, in general, characterized by the presence of a larger number of clumps than systems with no or low rotation. This effect, combined with the longer survival of substructure in rotating systems, leads to significant differences in the structural properties of systems with different initial rotation during the post-maximum contraction phase.

These differences are in part captured and illustrated by the time evolution of the fractal dimension shown in Fig. \ref{fig:fractal_comb}. 
This figure summarizes the structural evolution of the clusters we have studied, the initial growth, subsequent mergers, and disruption of clumps as the clusters evolve toward a smooth monolithic structure.
In the post-maximum-contraction phase, rapidly rotating systems are characterized by smaller values of the fractal dimension (an indication of a more clumpy structure) than systems with no or low rotation.
We point out that while the fractal dimension nicely captures the small scale substructure of clusters during their early evolution, it does not clearly reveal some significant large-scale structural differences in the clusters' final evolutionary stages. For example, at the end of the simulations all systems have very similar values of the fractal dimension, but the \URVH\ system is characterized by the presence of two subsystems, (see bottom right panel of Figs.  \ref{fig:clumpy_struct_xy} and \ref{fig:clumpy_struct_xz}) while in all the other systems starting with homogeneous density profile all clumps have merged and no significant substructure is present.

\subsection{Rotational properties}
\label{sec:rotation}
In this section we discuss the rotational properties at the end of the violent relaxation phase when the clusters have reached a virial equilibrium state.

In Fig. \ref{fig:Vcurve_low} we present the rotation curves for a few representative, initially rotating models.  The rotation curves are calculated from the rotational velocities of stars in concentric cylindrical radial bins that are aligned parallel to the rotation axis and display the median rotational velocity curve in the time interval 4.5-5.0 $t_{\rm ff}$ for \URLow, \CRLow, \URHigh, and \CRHigh, and 5.5-6.0 $t_{\rm ff}$ for \URVH\ and \CRVH, since they do not reach equilibrium until about $ 5.5 t_{\rm ff}$.

The rotation curves of all systems share a few general properties: they are all characterized by rotational velocities increasing with the distance from the cluster's centre to a maximum in the intermediate regions ($R_{\rm h}<R<3R_{\rm h}$), and decrease in the outer regions.

Initially homogeneous clusters have stronger rotation than the initially fractal counterparts; the substructure in the spatial distribution of initially fractal clusters and the various clump-clump interactions during the violent relaxation phase lead to shallower and less ordered collapse resulting in systems with a weaker rotation.

As shown in Fig. \ref{fig:Vcurve_low}, we find that, in several of the systems we have studied, massive stars tend to have rotational velocities around the $z$-axis of the cluster larger than low-mass stars. We interpret this trend as the outcome of the combined effect of angular momentum conservation, spatial segregation, and anisotropic segregation of massive stars (we will further discuss mass segregation in Sect. \ref{sect:MassSeg}). This effect is weaker in initially fractal clusters in which the less ordered nature of the violent relaxation collapse and re-expansion phases leads to a more effective kinematic randomization.

The more complex dynamics and clump-clump interactions occurring during the violent relaxation of initially fractal systems can produce an interesting kinematic feature in the inner rotation curve of slowly rotating systems; specifically, as shown in the right panel of Fig. \ref{fig:Vcurve_low}, we find that in the inner regions of the \CRLow\ system massive stars are counter-rotating relative to the rest of the cluster. This is one of the manifestations of the dynamical link between stellar mass and angular momentum orientation, which will be further discussed later in the paper (see Section \ref{sec:animassseg}).

Fig. \ref{fig:LzLEvo_all} shows the time evolution of the distribution of the orbital alignment to the $z$-axis, $L_{\rm z}/L$ (where $L$ is the magnitude of the angular momentum vector, and $L_{\rm z}$ is the component around the $z$-axis),
for the models \URHigh\ (left panel) and \CRHigh\ (right panel). 
The orbital alignment ranges from -1 (full orbital alignment with the -$z$-axis) to 1 (full orbital alignment with the +$z$-axis); for the rotating systems we have studied most stars have initially $L_{\rm z}/L > 0.5$. The two panels of Fig.  \ref{fig:LzLEvo_all} clearly show that, as the clusters evolve, the strong initial orbital alignment is  in part erased as stars are scattered by interactions with other stars and clumps present during these stages.  
The evolution of an initially fractal system (the \CRHigh\ system; right panel of Fig. \ref{fig:LzLEvo_all}), shows evidence of an inner counter-rotating subsystem in the early phases of evolution after the maximum contraction.  The strength and nature of the counter-rotating component in the initially fractal systems is subject to stochastic effects, and can be less evident as the initial rotation of the system is increased. In our \CRLow\ model, this counter-rotating component is more significant and lasts longer; as we will discuss later, massive stars tend to preferentially segregate on orbits aligned with the cluster bulk rotation and for the \CRLow\ simulation the inner negative rotational velocity of massive stars shown in the top right panel of Fig. \ref{fig:Vcurve_low} is a manifestation of the counter-rotating component.

In Fig. \ref{fig:ret_framexz_CRLow}, we further explore the evolution of the \CRLow\ system  and the dynamical origin of the stars populating the inner counter-rotating core by combining the information on the cluster's structural evolution with that of the distribution of $L_{\rm z}/L$.
Different panels show the evolution of stars separated according to the values of $L_{\rm z}/L$. As shown in this figure, as the cluster starts its collapse, a  population of counter-rotating stars forms as a result of the clump-clump interactions and scatterings among the various substructures already present in the very early evolutionary phases of this system. Strongly counter-rotating stars (with $L_{\rm z}/L < -0.75$) eventually settle down in a centrally concentrated and flattened core overlapping with the prograde population.

\subsection{Mass segregation and evolution towards energy equipartition}

\subsubsection{Mass segregation}
\label{sect:MassSeg}
Mass segregation is one of the manifestations of two-body relaxation in multi-mass collisional systems and is usually associated with the system's long-term evolution. However, as discussed in the Introduction, a number of studies have shown the development of significant mass segregation also during a cluster's early violent relaxation stages.

Here, we further explore this early segregation and its link with the evolution of the clusters' kinematic properties.

The two panels of Fig. \ref{fig:HMRadii} show the time evolution of the half-mass radius of stars in different mass bins for two representative models. As clearly illustrated by this figure, clusters develop mass segregation during the violent relaxation process and settle down to an equilibrium configuration in which the half-mass radius of the most massive stars is significantly smaller than that of the low-mass stars. A comparison between the two panels of Fig. \ref{fig:HMRadii} also shows that the initially homogeneous cluster settles into a more compact final equilibrium configuration. This is the consequence of the differences between the initial spatial distributions of the two systems: the initially homogeneous systems undergoes a more coherent and deeper collapse resulting in a more compact final configuration.
As for the effect of rotation on the final cluster size, we find that clusters initially rotating more rapidly are characterized by a larger final size. The effect of rotation on the final size is similar for both homogeneous and fractal systems but is weaker in the latter case.

A variety of different parameters have been introduced in the literature to quantify the degree of mass segregation (see e.g. \citealt{2015PaGo}, \citealt{2011MaCl}); here we are interested in quantifying the segregation developing on a local scale in the various substructures present during the cluster's early evolution and we use  the local density of stars in different mass bins to measure the degree of mass segregation. Fig. \ref{fig:MSeg} shows the time evolution of the median local number density (where the local number density is calculated by the distance to the 6th nearest-neighbour; \citealt{CasHut}) for stars in different mass bins and further illustrates the dynamical path leading to the mass segregation we find in all our systems. Studying the dependence of the local density on the stellar mass allows us to probe the development of mass segregation on a local scale during the early cluster's evolution when the systems are characterized by an irregular and clumpy spatial structure: stars with higher local number densities are, in general, those located in the central regions of the various clumps that are formed and destroyed during the violent relaxation process.

This figure clearly reveals the development of mass segregation during the very early stages of violent relaxation before the systems reaches its first maximum contraction (at $t/t_{\rm ff} \sim 1)$; massive stars preferentially migrate towards regions of high local density at the centre of the various clumps present in the evolving clusters. Memory of this early segregation is then preserved in the late stages of violent relaxation as the various clumps and substructures interact with each other and eventually merge resulting in a single monolithic cluster (see e.g. \citealt{2007McVe}).

The comparison between the evolution of the median local density for the initially homogeneous system and that of the initially fractal system shows that the homogeneous system develops a stronger mass segregation than the fractal system.
This could be the result of the stronger effects of two-body relaxation in the smaller and denser clumps forming during the evolution of the homogeneous system; in fact, while clumps are already present in the initial conditions of the fractal system, they are, in general, larger and less dense than those arising during the evolution of the initially homogeneous clusters (see Fig. \ref{fig:uni_struct_xy} and \ref{fig:uni_struct_xz}).

In Fig. \ref{fig:Mseg_comb}, we compare the time evolution of the median local density for the stars in the most massive bin for systems with different initial rotations and spatial distributions.  For all the values of the initial rotation we have investigated, all of our systems with an initially homogeneous spatial distribution (left panel) are characterized by a stronger mass segregation than initially fractal systems (right panel) during and after the violent relaxation phase.

To compare the final mass segregation across all models, Fig. \ref{fig:HMR_late} displays the ratio of the half-mass radius of stars in the most massive group (stars with $m> 20 m_{\odot}$) to the half-mass radius of all stars as a function of the strength of the initial rotation (as measured by the ratio $Q_{\rm rot}/Q_{\rm tidal}$). This figure shows that mass segregation is a common feature of the equilibrium configuration reached at the end of the violent relaxation stage by all the systems we have studied.
As shown in Fig. \ref{fig:HMR_late}, the final degree of mass segregation depends both on the cluster's initial spatial distribution and rotation: models starting with a homogeneous density profile are characterized by a stronger final segregation than those starting with a fractal spatial distribution, and models with high rotation are less  spatially segregated than those with slow initial rotation. 

Finally, Fig. \ref{fig:Dens_mass} shows the spherical density profiles for stars in four different mass bins, normalized to the density of each mass range in the innermost radial bin, calculated at the end of the simulations when the systems have reached an equilibrium state, from $4.5-5 t_{\rm ff}$.
This figure highlights the shape of the density profile in the inner regions of the cluster, and shows that more massive stars (in particular stars with m>20$M_{\sun}$) are more spatially segregated and have a more cuspy density profile than low-mass stars.

\subsubsection{Anisotropic mass segregation}
\label{sec:animassseg}
In this section, we show how the presence of rotation further enriches the early cluster dynamics and introduces a new fundamental link between early mass segregation and the kinematic properties of clusters emerging at the end of the violent relaxation phase.

Fig. \ref{fig:AvgMassEn} shows a map of the mean stellar mass on the $R$-$L_{\rm z}/L$ plane (where $R$ is the 2D distance of various Lagrangian shells from the cluster's centre). Each cell of this map is colour-coded according to $\ln(\epsilon)$, where $\epsilon$ is a parameter introduced in \cite{AnisotropicMS} and defined as the average mass of stars in that cell normalized to the average mass of all stars in the same radial shell:

\begin{equation}
    \epsilon(R,L_{\rm z}/L)=\frac{\overline{m}(R,L_{\rm z}/L)}{\overline{m}(R)}.
    \label{eq:mass_en}
\end{equation}

This figure clearly shows evidence of anisotropic mass segregation: more massive stars tend to kinematically segregate towards larger values of $L_{\rm z}/L$ leading, at all radii, to larger average mean stellar masses in the cells corresponding to larger values of $L_{\rm z}/L$. We find this trend in all of the rotating models studied regardless of the initial spatial distribution, although, as shown in Fig. \ref{fig:AvgMassEn}, the trend is stronger for initially homogeneous models.

\cite{AnisotropicMS} have recently explored the long-term evolution of rotating multi-mass globular clusters and found that, in addition to the well known radial segregation, massive stars tend to segregate into orbits with angular momentum aligned with the cluster's angular momentum. Our results show that the same kinematic segregation occurs during early evolutionary phases as a cluster undergoes violent relaxation and evolves toward an equilibrium configuration.

In order to further explore the kinematic mass segregation and its link with the orbital inclination, we introduce a parameter $\tilde\epsilon$, defined as the integral of $\epsilon$ over the inner regions of the cluster (a similar metric was introduced in \cite{AnisotropicMS}): 

\begin{equation}
    \tilde\epsilon(L_{\rm z}/L)=\epsilon(R<R_{25\%},L_{\rm z}/L)
    \label{eq:mass_en_eff}
\end{equation}

\noindent where $R_{25\%}$ is the radius containing the inner 25\% of stars.

Fig. \ref{fig:AvgMassEn_table} visualizes $\tilde\epsilon$ as a function of $L_{\rm z}/L$ in each cluster, and shows evidence of anisotropic mass segregation in the rotating models, with a stronger degree in the initially homogeneous models (left panel of Fig. \ref{fig:AvgMassEn_table}).

\subsubsection{Evolution towards energy equipartition}
One of the consequences of the effects of two-body relaxation is the evolution towards energy equipartition.

The evolution towards equipartition has been extensively studied in the context of the long-term evolution of star clusters; several studies have shown that full energy equipartition cannot be achieved in systems with a realistic mass function (see e.g. \citealt{NoEnEq}, \citealt{eneq_webb})  and explored the degree of partial energy equipartition reached by clusters as function of the distance from the cluster's centre and the cluster's dynamical phase (\citealt{meq}, \citealt{eneq_webb}, \citealt{2018BiWeSi}, \citealt{2018LiBevdM}, \citealt{2021PaVe})
The extent to which early mass segregation developing during  the violent relaxation phase is accompanied by the kinematic evolution towards energy equipartition, on the other hand, has received relatively little attention.

In Fig. \ref{fig:DispvsR}, we show the final radial profiles of the projected velocity dispersion (calculated along a line of sight parallel to the cluster's rotation axis) for stars in different mass groups for \URHigh\ (left panel) and \CRHigh\ (right panel). In both cases our results show clear evidence of evolution towards energy equipartition in the clusters' inner regions ($R<R_{\rm h}$) where more massive stars have smaller velocity dispersion than low-mass stars.
This effect is more pronounced in the initially homogeneous system, but is clearly present and significant in the initially fractal system as well.

In order to compare the degree of evolution towards energy equipartition in all the models studied, Fig. \ref{fig:EEqui} plots the velocity dispersion along the $z$-axis, $\sigma_{\rm z}$, versus stellar mass for the innermost 10\% of the stars in the cluster ($R\lesssim 0.3 R_{\rm h}$). This figure shows that all models are characterized by a dependence of the velocity dispersion on the stellar mass consistent with that expected for systems evolving towards energy equipartition. 
In order to further quantify the degree of evolution towards energy equipartition, we have fit these profiles with the exponential function introduced by \cite{meq}:

\begin{equation}
\label{eq:meq_sig}
    \sigma(m) =\left\{\begin{array}{cc}
    \sigma_0 \exp\left(-\frac{1}{2} \frac{m}{m_{\rm eq}}\right) & m<m_{\rm eq}\\
    \sigma_0 \left(\frac{m}{m_{\rm eq}}\right)^{-1/2} & m\geq m_{\rm eq}\\
         \end{array}  \right.
\end{equation}

\noindent where $m_{\rm eq}$, the {\it equipartition mass}, and $\sigma_0$ are the free parameters. 

Fig. \ref{fig:EEqui_mass} plots the equipartition mass of each cluster's central region and shows clear evidence of evolution towards energy equipartition for all clusters. For models with slow or moderate rotation, initially homogeneous models have smaller values of $m_{\rm eq}$, indicating a stronger degree of evolution towards energy equipartition. We do not find any significant trend between initial rotation and the final degree of evolution towards energy equipartition.

Quantifying the degree of evolution towards energy equipartition during the violent relaxation phase is not straightforward, since the system is still not in equilibrium and is characterized by complex structural and kinematic properties.
In order to further explore the evolution towards energy equipartition and trace the dynamical path leading to the mass-dependent velocity dispersion found when our models reach virial equilibrium, we have adopted the following method.
We have compared the speed, $s$, of each star in the cluster with the median speed of its 20 nearest-neighbours, $s_{\rm NN}$. In a system in which the velocity of stars is independent of the stellar mass, the fraction of stars with speeds smaller than the median of its nearest neighbours is expected to be about 50\% for any mass group. For clusters evolving towards energy equipartition, massive stars will tend to have smaller velocities than low-mass stars, and the fraction of massive stars with speeds smaller than the median speed of their neighbours is therefore expected to increase. 

Fig. \ref{fig:EEqui_localall} (top panels) shows the time evolution of the fraction of stars with speeds smaller than the median speed of their 20 nearest-neighbours: as expected, the fraction is initially equal to about 50\% for stars in all the mass groups but, as the system starts to evolve, it immediately increases for the most massive groups  clearly showing evidence of the early evolution towards a mass-dependent kinematics.

This result combined with that of the evolution of spatial segregation in the bottom panels of Fig. \ref{fig:EEqui_localall}, where this analysis is carried out only for the stars with the top 10\% of the local number density, clearly reveals the dynamical path behind the development of the structural and kinematic fingerprints associated with the collisional effects acting during the cluster early dynamical phases.

Finally, in order to further quantify the early evolution towards energy equipartition, we show in Fig. \ref{fig:EEqui_div_localall} the time evolution of the median of $s/s_{\rm NN}$.
This figure provides a more quantitative measure of the extent of the differences between the speeds of stars in different mass bins and those of their nearest-neighbours.

\section{Conclusions}
\label{Conclusions}
In this paper, we have explored the early evolution of rotating multi-mass star clusters as they undergo the violent relaxation phase in the presence of a weak external tidal field and evolve towards an equilibrium state. We have focused our attention, in particular, on the effects due to the clusters' initial rotation and the dependence of the evolving structural and kinematic properties on the stellar mass.

The main conclusions of our study are the following.
\begin{itemize}
\item We have followed the evolution of the structural properties of clusters and found that, regardless of whether a cluster starts with a homogeneous or fractal spatial distribution, during the violent relaxation phase, clusters develop a clumpy distribution; individual clumps grow, interact with each other, and eventually merge producing a single monolithic system. Rotation plays a key role in this phase as it delays the clump merging and extends the phase during which a cluster is characterized by a clumpy structure.
\item The final equilibrium state in all of the systems we have studied are characterized by a rotation curve that increases with the distance from the cluster’s centre to a maximum in the intermediate regions ($R_{\rm h}<R<3R_{\rm h}$), and decreases in the outer regions. Our study has revealed a dependence of the final rotation velocity on the star mass: high-mass stars tend to rotate more rapidly. This trend is stronger in systems starting with a homogeneous spatial distribution.
\item We find that systems starting with an initial fractal distribution with low/moderate initial rotation can be characterized by the presence of an inner counter-rotating subsystem.
\item All the systems studied evolve into a final equilibrium in which massive stars are segregated in the cluster inner regions. We show that massive stars tend to segregate towards the inner region of the various clumps present already in the very early phases of violent relaxation, and such segregation is then preserved in the final monolithic cluster emerging at the end of violent relaxation.
Mass segregation is stronger in initially homogeneous systems and in systems with low initial rotation.
\item We show that mass segregation is accompanied by an evolution towards (partial) energy equipartition starting during the very early stages of violent relaxation. The final equilibrium states are characterized by a velocity dispersion decreasing for increasing stellar masses (in particular in the cluster inner regions) as expected for stellar systems evolving towards energy equipartition. The final degree of equipartition is slightly stronger for systems starting with a homogeneous spatial distribution.
\item   In addition to spatially segregating in the cluster inner regions, massive stars show evidence of kinematic segregation as they preferentially segregate into orbits with angular momentum aligned with the cluster’s angular momentum.
This effect was previously found by \cite{AnisotropicMS} in a study of the long-term evolution of multi-mass rotating star clusters driven by two-body relaxation. Our analysis shows that a similar anisotropic segregation also arises during the cluster's early violent relaxation phase.
  The development of this kinematic segregation, along with mass segregation and evolution towards energy equipartition, provides another example of the manifestation of collisional effects during the violent relaxation and early evolution of star clusters.
  \end{itemize}
In future studies we will further expand the analysis presented here by including the effects of stellar evolution, by considering a broader range of structural and kinematic initial properties, testing different tidal field strengths, and we will extend our study to explore the long-term evolution of  multi-mass rotating clusters.

\section*{Acknowledgements}
This research was supported in part by Lilly Endowment, Inc., through its support for the Indiana University Pervasive Technology Institute.
ALV acknowledges support from a UKRI Future Leaders Fellowship (MR/S018859/1).
We thank the referee for a careful review of the manuscript and insightful comments that helped us to clarify the presentation of our results.

\section*{Data availability statement}
The data presented in this article may be shared on reasonable request to the corresponding author.

\bibliographystyle{mnras}
\bibliography{example}

\begin{thebibliography}{}
\makeatletter
\relax
\def\mn@urlcharsother{\let\do\@makeother \do\$\do\&\do\#\do\^\do\_\do\%\do\~}
\def\mn@doi{\begingroup\mn@urlcharsother \@ifnextchar [ {\mn@doi@}
  {\mn@doi@[]}}
\def\mn@doi@[#1]#2{\def\@tempa{#1}\ifx\@tempa\@empty \href
  {http://dx.doi.org/#2} {doi:#2}\else \href {http://dx.doi.org/#2} {#1}\fi
  \endgroup}
\def\mn@eprint#1#2{\mn@eprint@#1:#2::\@nil}
\def\mn@eprint@arXiv#1{\href {http://arxiv.org/abs/#1} {{\tt arXiv:#1}}}
\def\mn@eprint@dblp#1{\href {http://dblp.uni-trier.de/rec/bibtex/#1.xml}
  {dblp:#1}}
\def\mn@eprint@#1:#2:#3:#4\@nil{\def\@tempa {#1}\def\@tempb {#2}\def\@tempc
  {#3}\ifx \@tempc \@empty \let \@tempc \@tempb \let \@tempb \@tempa \fi \ifx
  \@tempb \@empty \def\@tempb {arXiv}\fi \@ifundefined
  {mn@eprint@\@tempb}{\@tempb:\@tempc}{\expandafter \expandafter \csname
  mn@eprint@\@tempb\endcsname \expandafter{\@tempc}}}

\bibitem[\protect\citeauthoryear{{Aarseth}}{{Aarseth}}{2003}]{nbody6_2003}
{Aarseth} S.~J.,  2003, {Gravitational N-Body Simulations}

\bibitem[\protect\citeauthoryear{{Aarseth}, {Lin}  \& {Papaloizou}}{{Aarseth}
  et~al.}{1988}]{1988AaLi}
{Aarseth} S.~J.,  {Lin} D.~N.~C.,   {Papaloizou} J.~C.~B.,  1988, \mn@doi
  [\apj] {10.1086/165895}, \href
  {https://ui.adsabs.harvard.edu/abs/1988ApJ...324..288A} {324, 288}

\bibitem[\protect\citeauthoryear{{Allison}, {Goodwin}, {Parker}, {Portegies
  Zwart}, {de Grijs}  \& {Kouwenhoven}}{{Allison} et~al.}{2009}]{2009AlGo}
{Allison} R.~J.,  {Goodwin} S.~P.,  {Parker} R.~J.,  {Portegies Zwart} S.~F.,
  {de Grijs} R.,   {Kouwenhoven} M.~B.~N.,  2009, \mn@doi [\mnras]
  {10.1111/j.1365-2966.2009.14508.x}, \href
  {https://ui.adsabs.harvard.edu/abs/2009MNRAS.395.1449A} {395, 1449}

\bibitem[\protect\citeauthoryear{{Allison}, {Goodwin}, {Parker}, {Portegies
  Zwart}  \& {de Grijs}}{{Allison} et~al.}{2010}]{2010AlGo}
{Allison} R.~J.,  {Goodwin} S.~P.,  {Parker} R.~J.,  {Portegies Zwart} S.~F.,
  {de Grijs} R.,  2010, \mn@doi [\mnras] {10.1111/j.1365-2966.2010.16939.x},
  \href {https://ui.adsabs.harvard.edu/abs/2010MNRAS.407.1098A} {407, 1098}

\bibitem[\protect\citeauthoryear{Baddeley, Rubak  \& Turner}{Baddeley
  et~al.}{2015}]{spatstatBook}
Baddeley A.,  Rubak E.,   Turner R.,  2015, Spatial Point Patterns: Methodology
  and Applications with {R}.
Chapman and Hall/CRC, London, \url
  {http://www.crcpress.com/books/details/9781482210200/}

\bibitem[\protect\citeauthoryear{{Ballone}, {Mapelli}, {Di Carlo},
  {Torniamenti}, {Spera}  \& {Rastello}}{{Ballone} et~al.}{2020}]{2020BaMa}
{Ballone} A.,  {Mapelli} M.,  {Di Carlo} U.~N.,  {Torniamenti} S.,  {Spera} M.,
    {Rastello} S.,  2020, \mn@doi [\mnras] {10.1093/mnras/staa1383}, \href
  {https://ui.adsabs.harvard.edu/abs/2020MNRAS.496...49B} {496, 49}

\bibitem[\protect\citeauthoryear{{Ballone}, {Torniamenti}, {Mapelli}, {Di
  Carlo}, {Spera}, {Rastello}, {Gaspari}  \& {Iorio}}{{Ballone}
  et~al.}{2021}]{2021BaTo}
{Ballone} A.,  {Torniamenti} S.,  {Mapelli} M.,  {Di Carlo} U.~N.,  {Spera} M.,
   {Rastello} S.,  {Gaspari} N.,   {Iorio} G.,  2021, \mn@doi [\mnras]
  {10.1093/mnras/staa3763}, \href
  {https://ui.adsabs.harvard.edu/abs/2021MNRAS.501.2920B} {501, 2920}

\bibitem[\protect\citeauthoryear{{Banerjee} \& {Kroupa}}{{Banerjee} \&
  {Kroupa}}{2014}]{2014BaKr}
{Banerjee} S.,  {Kroupa} P.,  2014, \mn@doi [\apj]
  {10.1088/0004-637X/787/2/158}, \href
  {https://ui.adsabs.harvard.edu/abs/2014ApJ...787..158B} {787, 158}

\bibitem[\protect\citeauthoryear{{Banerjee} \& {Kroupa}}{{Banerjee} \&
  {Kroupa}}{2017}]{2017BaKr}
{Banerjee} S.,  {Kroupa} P.,  2017, \mn@doi [\aap]
  {10.1051/0004-6361/201526928}, \href
  {https://ui.adsabs.harvard.edu/abs/2017A&A...597A..28B} {597, A28}

\bibitem[\protect\citeauthoryear{{Banerjee} \& {Kroupa}}{{Banerjee} \&
  {Kroupa}}{2018}]{2018BaKr}
{Banerjee} S.,  {Kroupa} P.,  2018, {Formation of Very Young Massive Clusters
  and Implications for Globular Clusters}.
p.~143, \mn@doi{10.1007/978-3-319-22801-3_6}

\bibitem[\protect\citeauthoryear{{Baumgardt} \& {Hilker}}{{Baumgardt} \&
  {Hilker}}{2018}]{2018BaHi}
{Baumgardt} H.,  {Hilker} M.,  2018, \mn@doi [\mnras] {10.1093/mnras/sty1057},
  \href {https://ui.adsabs.harvard.edu/abs/2018MNRAS.478.1520B} {478, 1520}

\bibitem[\protect\citeauthoryear{{Bellini}, {Bianchini}, {Varri}, {Anderson},
  {Piotto}, {van der Marel}, {Vesperini}  \& {Watkins}}{{Bellini}
  et~al.}{2017}]{2017BeBi}
{Bellini} A.,  {Bianchini} P.,  {Varri} A.~L.,  {Anderson} J.,  {Piotto} G.,
  {van der Marel} R.~P.,  {Vesperini} E.,   {Watkins} L.~L.,  2017, \mn@doi
  [\apj] {10.3847/1538-4357/aa7c5f}, \href
  {https://ui.adsabs.harvard.edu/abs/2017ApJ...844..167B} {844, 167}

\bibitem[\protect\citeauthoryear{{Bianchini}, {van de Ven}, {Norris},
  {Schinnerer}  \& {Varri}}{{Bianchini} et~al.}{2016}]{meq}
{Bianchini} P.,  {van de Ven} G.,  {Norris} M.~A.,  {Schinnerer} E.,   {Varri}
  A.~L.,  2016, \mn@doi [\mnras] {10.1093/mnras/stw552}, \href
  {https://ui.adsabs.harvard.edu/abs/2016MNRAS.458.3644B} {458, 3644}

\bibitem[\protect\citeauthoryear{{Bianchini}, {Webb}, {Sills}  \&
  {Vesperini}}{{Bianchini} et~al.}{2018a}]{2018BiWeSi}
{Bianchini} P.,  {Webb} J.~J.,  {Sills} A.,   {Vesperini} E.,  2018a, \mn@doi
  [\mnras] {10.1093/mnrasl/sly013}, \href
  {https://ui.adsabs.harvard.edu/abs/2018MNRAS.475L..96B} {475, L96}

\bibitem[\protect\citeauthoryear{{Bianchini}, {van der Marel}, {del Pino},
  {Watkins}, {Bellini}, {Fardal}, {Libralato}  \& {Sills}}{{Bianchini}
  et~al.}{2018b}]{2018Biva}
{Bianchini} P.,  {van der Marel} R.~P.,  {del Pino} A.,  {Watkins} L.~L.,
  {Bellini} A.,  {Fardal} M.~A.,  {Libralato} M.,   {Sills} A.,  2018b, \mn@doi
  [\mnras] {10.1093/mnras/sty2365}, \href
  {https://ui.adsabs.harvard.edu/abs/2018MNRAS.481.2125B} {481, 2125}

\bibitem[\protect\citeauthoryear{{Casertano} \& {Hut}}{{Casertano} \&
  {Hut}}{1985}]{CasHut}
{Casertano} S.,  {Hut} P.,  1985, \mn@doi [\apj] {10.1086/163589}, \href
  {https://ui.adsabs.harvard.edu/abs/1985ApJ...298...80C} {298, 80}

\bibitem[\protect\citeauthoryear{{Cohen}, {Bellini}, {Libralato}, {Correnti},
  {Brown}  \& {Kalirai}}{{Cohen} et~al.}{2021}]{2021CoBe}
{Cohen} R.~E.,  {Bellini} A.,  {Libralato} M.,  {Correnti} M.,  {Brown} T.~M.,
   {Kalirai} J.~S.,  2021, \mn@doi [\aj] {10.3847/1538-3881/abd036}, \href
  {https://ui.adsabs.harvard.edu/abs/2021AJ....161...41C} {161, 41}

\bibitem[\protect\citeauthoryear{{Cordoni}, {Milone}, {Mastrobuono-Battisti},
  {Marino}, {Lagioia}, {Tailo}, {Baumgardt}  \& {Hilker}}{{Cordoni}
  et~al.}{2020}]{2020CoMi}
{Cordoni} G.,  {Milone} A.~P.,  {Mastrobuono-Battisti} A.,  {Marino} A.~F.,
  {Lagioia} E.~P.,  {Tailo} M.,  {Baumgardt} H.,   {Hilker} M.,  2020, \mn@doi
  [\apj] {10.3847/1538-4357/ab5aee}, \href
  {https://ui.adsabs.harvard.edu/abs/2020ApJ...889...18C} {889, 18}

\bibitem[\protect\citeauthoryear{{Daffern-Powell} \& {Parker}}{{Daffern-Powell}
  \& {Parker}}{2020}]{2020DaPa}
{Daffern-Powell} E.~C.,  {Parker} R.~J.,  2020, \mn@doi [\mnras]
  {10.1093/mnras/staa575}, \href
  {https://ui.adsabs.harvard.edu/abs/2020MNRAS.493.4925D} {493, 4925}

\bibitem[\protect\citeauthoryear{{Dalessandro}, {Raso}, {Kamann}, {Bellazzini},
  {Vesperini}, {Bellini}  \& {Beccari}}{{Dalessandro} et~al.}{2021a}]{2021DaRa}
{Dalessandro} E.,  {Raso} S.,  {Kamann} S.,  {Bellazzini} M.,  {Vesperini} E.,
  {Bellini} A.,   {Beccari} G.,  2021a, \mn@doi [\mnras]
  {10.1093/mnras/stab1257}, \href
  {https://ui.adsabs.harvard.edu/abs/2021MNRAS.tmp.1236D} {}

\bibitem[\protect\citeauthoryear{{Dalessandro} et~al.,}{{Dalessandro}
  et~al.}{2021b}]{2021DaVa}
{Dalessandro} E.,  et~al., 2021b, \mn@doi [\apj] {10.3847/1538-4357/abda43},
  \href {https://ui.adsabs.harvard.edu/abs/2021ApJ...909...90D} {909, 90}

\bibitem[\protect\citeauthoryear{{Dom{\'\i}nguez}, {Fellhauer}, {Bla{\~n}a},
  {Farias}  \& {Dabringhausen}}{{Dom{\'\i}nguez} et~al.}{2017}]{2017DoFe}
{Dom{\'\i}nguez} R.,  {Fellhauer} M.,  {Bla{\~n}a} M.,  {Farias} J.~P.,
  {Dabringhausen} J.,  2017, \mn@doi [\mnras] {10.1093/mnras/stx1883}, \href
  {https://ui.adsabs.harvard.edu/abs/2017MNRAS.472..465D} {472, 465}

\bibitem[\protect\citeauthoryear{{Einsel} \& {Spurzem}}{{Einsel} \&
  {Spurzem}}{1999}]{1999EiSp}
{Einsel} C.,  {Spurzem} R.,  1999, \mn@doi [\mnras]
  {10.1046/j.1365-8711.1999.02083.x}, \href
  {https://ui.adsabs.harvard.edu/abs/1999MNRAS.302...81E} {302, 81}

\bibitem[\protect\citeauthoryear{{Ernst}, {Glaschke}, {Fiestas}, {Just}  \&
  {Spurzem}}{{Ernst} et~al.}{2007}]{2007ErGl}
{Ernst} A.,  {Glaschke} P.,  {Fiestas} J.,  {Just} A.,   {Spurzem} R.,  2007,
  \mn@doi [\mnras] {10.1111/j.1365-2966.2007.11602.x}, \href
  {https://ui.adsabs.harvard.edu/abs/2007MNRAS.377..465E} {377, 465}

\bibitem[\protect\citeauthoryear{{Fabricius} et~al.,}{{Fabricius}
  et~al.}{2014}]{2014FaMa}
{Fabricius} M.~H.,  et~al., 2014, \mn@doi [\apjl]
  {10.1088/2041-8205/787/2/L26}, \href
  {https://ui.adsabs.harvard.edu/abs/2014ApJ...787L..26F} {787, L26}

\bibitem[\protect\citeauthoryear{{Ferraro} et~al.,}{{Ferraro}
  et~al.}{2018}]{2018FeMu}
{Ferraro} F.~R.,  et~al., 2018, \mn@doi [\apj] {10.3847/1538-4357/aabe2f},
  \href {https://ui.adsabs.harvard.edu/abs/2018ApJ...860...50F} {860, 50}

\bibitem[\protect\citeauthoryear{{Fujii} \& {Portegies Zwart}}{{Fujii} \&
  {Portegies Zwart}}{2016}]{2016FuPo}
{Fujii} M.~S.,  {Portegies Zwart} S.,  2016, \mn@doi [\apj]
  {10.3847/0004-637X/817/1/4}, \href
  {https://ui.adsabs.harvard.edu/abs/2016ApJ...817....4F} {817, 4}

\bibitem[\protect\citeauthoryear{{Fujii}, {Saitoh}  \& {Portegies
  Zwart}}{{Fujii} et~al.}{2012}]{2012FuSa}
{Fujii} M.~S.,  {Saitoh} T.~R.,   {Portegies Zwart} S.~F.,  2012, \mn@doi
  [\apj] {10.1088/0004-637X/753/1/85}, \href
  {https://ui.adsabs.harvard.edu/abs/2012ApJ...753...85F} {753, 85}

\bibitem[\protect\citeauthoryear{{Getman}, {Kuhn}, {Feigelson}, {Broos}, {Bate}
   \& {Garmire}}{{Getman} et~al.}{2018}]{2018mystix}
{Getman} K.~V.,  {Kuhn} M.~A.,  {Feigelson} E.~D.,  {Broos} P.~S.,  {Bate}
  M.~R.,   {Garmire} G.~P.,  2018, \mn@doi [\mnras] {10.1093/mnras/sty473},
  \href {https://ui.adsabs.harvard.edu/abs/2018MNRAS.477..298G} {477, 298}

\bibitem[\protect\citeauthoryear{{Getman}, {Feigelson}, {Kuhn}  \&
  {Garmire}}{{Getman} et~al.}{2019}]{2019GeFe}
{Getman} K.~V.,  {Feigelson} E.~D.,  {Kuhn} M.~A.,   {Garmire} G.~P.,  2019,
  \mn@doi [\mnras] {10.1093/mnras/stz1457}, \href
  {https://ui.adsabs.harvard.edu/abs/2019MNRAS.487.2977G} {487, 2977}

\bibitem[\protect\citeauthoryear{{Goodwin} \& {Whitworth}}{{Goodwin} \&
  {Whitworth}}{2004}]{2004GoWh}
{Goodwin} S.~P.,  {Whitworth} A.~P.,  2004, \mn@doi [\aap]
  {10.1051/0004-6361:20031529}, \href
  {https://ui.adsabs.harvard.edu/abs/2004A&A...413..929G} {413, 929}

\bibitem[\protect\citeauthoryear{{Hong}, {Kim}, {Lee}  \& {Spurzem}}{{Hong}
  et~al.}{2013}]{2013HoKi}
{Hong} J.,  {Kim} E.,  {Lee} H.~M.,   {Spurzem} R.,  2013, \mn@doi [\mnras]
  {10.1093/mnras/stt099}, \href
  {https://ui.adsabs.harvard.edu/abs/2013MNRAS.430.2960H} {430, 2960}

\bibitem[\protect\citeauthoryear{{Jindal}, {Webb}  \& {Bovy}}{{Jindal}
  et~al.}{2019}]{2019JiWe}
{Jindal} A.,  {Webb} J.~J.,   {Bovy} J.,  2019, \mn@doi [\mnras]
  {10.1093/mnras/stz1586}, \href
  {https://ui.adsabs.harvard.edu/abs/2019MNRAS.487.3693J} {487, 3693}

\bibitem[\protect\citeauthoryear{{Kamann} et~al.,}{{Kamann}
  et~al.}{2018}]{2018KaHu}
{Kamann} S.,  et~al., 2018, \mn@doi [\mnras] {10.1093/mnras/stx2719}, \href
  {https://ui.adsabs.harvard.edu/abs/2018MNRAS.473.5591K} {473, 5591}

\bibitem[\protect\citeauthoryear{{Kroupa}}{{Kroupa}}{2001}]{KIMF}
{Kroupa} P.,  2001, \mn@doi [\mnras] {10.1046/j.1365-8711.2001.04022.x}, \href
  {https://ui.adsabs.harvard.edu/abs/2001MNRAS.322..231K} {322, 231}

\bibitem[\protect\citeauthoryear{{Kuhn}, {Povich}, {Luhman}, {Getman}, {Busk}
  \& {Feigelson}}{{Kuhn} et~al.}{2013}]{2014mystix}
{Kuhn} M.~A.,  {Povich} M.~S.,  {Luhman} K.~L.,  {Getman} K.~V.,  {Busk} H.~A.,
    {Feigelson} E.~D.,  2013, \mn@doi [\apjs] {10.1088/0067-0049/209/2/29},
  \href {https://ui.adsabs.harvard.edu/abs/2013ApJS..209...29K} {209, 29}

\bibitem[\protect\citeauthoryear{{Kuhn} et~al.,}{{Kuhn}
  et~al.}{2014}]{2015mystixa}
{Kuhn} M.~A.,  et~al., 2014, \mn@doi [\apj] {10.1088/0004-637X/787/2/107},
  \href {https://ui.adsabs.harvard.edu/abs/2014ApJ...787..107K} {787, 107}

\bibitem[\protect\citeauthoryear{{Kuhn}, {Getman}  \& {Feigelson}}{{Kuhn}
  et~al.}{2015a}]{2015mystixb}
{Kuhn} M.~A.,  {Getman} K.~V.,   {Feigelson} E.~D.,  2015a, \mn@doi [\apj]
  {10.1088/0004-637X/802/1/60}, \href
  {https://ui.adsabs.harvard.edu/abs/2015ApJ...802...60K} {802, 60}

\bibitem[\protect\citeauthoryear{{Kuhn}, {Feigelson}, {Getman}, {Sills}, {Bate}
   \& {Borissova}}{{Kuhn} et~al.}{2015b}]{2017mystix}
{Kuhn} M.~A.,  {Feigelson} E.~D.,  {Getman} K.~V.,  {Sills} A.,  {Bate} M.~R.,
   {Borissova} J.,  2015b, \mn@doi [\apj] {10.1088/0004-637X/812/2/131}, \href
  {https://ui.adsabs.harvard.edu/abs/2015ApJ...812..131K} {812, 131}

\bibitem[\protect\citeauthoryear{{Kuhn}, {Hillenbrand}, {Sills}, {Feigelson}
  \& {Getman}}{{Kuhn} et~al.}{2019}]{2019KuHi}
{Kuhn} M.~A.,  {Hillenbrand} L.~A.,  {Sills} A.,  {Feigelson} E.~D.,   {Getman}
  K.~V.,  2019, \mn@doi [\apj] {10.3847/1538-4357/aaef8c}, \href
  {https://ui.adsabs.harvard.edu/abs/2019ApJ...870...32K} {870, 32}

\bibitem[\protect\citeauthoryear{{K{\"u}pper}, {Maschberger}, {Kroupa}  \&
  {Baumgardt}}{{K{\"u}pper} et~al.}{2011}]{mcluster}
{K{\"u}pper} A. H.~W.,  {Maschberger} T.,  {Kroupa} P.,   {Baumgardt} H.,
  2011, \mn@doi [\mnras] {10.1111/j.1365-2966.2011.19412.x}, \href
  {https://ui.adsabs.harvard.edu/abs/2011MNRAS.417.2300K} {417, 2300}

\bibitem[\protect\citeauthoryear{{Lanzoni} et~al.,}{{Lanzoni}
  et~al.}{2018}]{2018LaFe}
{Lanzoni} B.,  et~al., 2018, \mn@doi [\apj] {10.3847/1538-4357/aad810}, \href
  {https://ui.adsabs.harvard.edu/abs/2018ApJ...865...11L} {865, 11}

\bibitem[\protect\citeauthoryear{{Libralato} et~al.,}{{Libralato}
  et~al.}{2018a}]{2018LIBe}
{Libralato} M.,  et~al., 2018a, \mn@doi [\apj] {10.3847/1538-4357/aac6c0},
  \href {https://ui.adsabs.harvard.edu/abs/2018ApJ...861...99L} {861, 99}

\bibitem[\protect\citeauthoryear{{Libralato} et~al.,}{{Libralato}
  et~al.}{2018b}]{2018LiBevdM}
{Libralato} M.,  et~al., 2018b, \mn@doi [\apj] {10.3847/1538-4357/aac6c0},
  \href {https://ui.adsabs.harvard.edu/abs/2018ApJ...861...99L} {861, 99}

\bibitem[\protect\citeauthoryear{{Lim}, {Hong}, {Yun}, {Hwang}, {Kim}, {Lee},
  {Park}  \& {Park}}{{Lim} et~al.}{2020}]{2020LiHo}
{Lim} B.,  {Hong} J.,  {Yun} H.-S.,  {Hwang} N.,  {Kim} J.~S.,  {Lee} J.-E.,
  {Park} B.-G.,   {Park} S.,  2020, \mn@doi [\apj] {10.3847/1538-4357/aba0a3},
  \href {https://ui.adsabs.harvard.edu/abs/2020ApJ...899..121L} {899, 121}

\bibitem[\protect\citeauthoryear{{Mapelli}}{{Mapelli}}{2017}]{MapMi}
{Mapelli} M.,  2017, \mn@doi [\mnras] {10.1093/mnras/stx304}, \href
  {https://ui.adsabs.harvard.edu/abs/2017MNRAS.467.3255M} {467, 3255}

\bibitem[\protect\citeauthoryear{{Maschberger} \& {Clarke}}{{Maschberger} \&
  {Clarke}}{2011}]{2011MaCl}
{Maschberger} T.,  {Clarke} C.~J.,  2011, \mn@doi [\mnras]
  {10.1111/j.1365-2966.2011.19067.x}, \href
  {https://ui.adsabs.harvard.edu/abs/2011MNRAS.416..541M} {416, 541}

\bibitem[\protect\citeauthoryear{{McMillan}, {Vesperini}  \& {Portegies
  Zwart}}{{McMillan} et~al.}{2007}]{2007McVe}
{McMillan} S. L.~W.,  {Vesperini} E.,   {Portegies Zwart} S.~F.,  2007, \mn@doi
  [\apjl] {10.1086/511763}, \href
  {https://ui.adsabs.harvard.edu/abs/2007ApJ...655L..45M} {655, L45}

\bibitem[\protect\citeauthoryear{{Moeckel} \& {Bonnell}}{{Moeckel} \&
  {Bonnell}}{2009}]{2009MoBo}
{Moeckel} N.,  {Bonnell} I.~A.,  2009, \mn@doi [\mnras]
  {10.1111/j.1365-2966.2009.15499.x}, \href
  {https://ui.adsabs.harvard.edu/abs/2009MNRAS.400..657M} {400, 657}

\bibitem[\protect\citeauthoryear{{Nitadori} \& {Aarseth}}{{Nitadori} \&
  {Aarseth}}{2012}]{GPU_nbody6}
{Nitadori} K.,  {Aarseth} S.~J.,  2012, \mn@doi [\mnras]
  {10.1111/j.1365-2966.2012.21227.x}, \href
  {https://ui.adsabs.harvard.edu/abs/2012MNRAS.424..545N} {424, 545}

\bibitem[\protect\citeauthoryear{{Parker} \& {Goodwin}}{{Parker} \&
  {Goodwin}}{2015}]{2015PaGo}
{Parker} R.~J.,  {Goodwin} S.~P.,  2015, \mn@doi [\mnras]
  {10.1093/mnras/stv539}, \href
  {https://ui.adsabs.harvard.edu/abs/2015MNRAS.449.3381P} {449, 3381}

\bibitem[\protect\citeauthoryear{{Parker}, {Goodwin}, {Wright}, {Meyer}  \&
  {Quanz}}{{Parker} et~al.}{2016}]{2016PaGo}
{Parker} R.~J.,  {Goodwin} S.~P.,  {Wright} N.~J.,  {Meyer} M.~R.,   {Quanz}
  S.~P.,  2016, \mn@doi [\mnras] {10.1093/mnrasl/slw061}, \href
  {https://ui.adsabs.harvard.edu/abs/2016MNRAS.459L.119P} {459, L119}

\bibitem[\protect\citeauthoryear{{Pavl{\'\i}k} \& {Vesperini}}{{Pavl{\'\i}k} \&
  {Vesperini}}{2021}]{2021PaVe}
{Pavl{\'\i}k} V.,  {Vesperini} E.,  2021, \mn@doi [\mnras]
  {10.1093/mnrasl/slab026}, \href
  {https://ui.adsabs.harvard.edu/abs/2021MNRAS.504L..12P} {504, L12}

\bibitem[\protect\citeauthoryear{{Sills}, {Rieder}, {Scora}, {McCloskey}  \&
  {Jaffa}}{{Sills} et~al.}{2018}]{2018SiRi}
{Sills} A.,  {Rieder} S.,  {Scora} J.,  {McCloskey} J.,   {Jaffa} S.,  2018,
  \mn@doi [\mnras] {10.1093/mnras/sty681}, \href
  {https://ui.adsabs.harvard.edu/abs/2018MNRAS.477.1903S} {477, 1903}

\bibitem[\protect\citeauthoryear{{Swiggum} et~al.,}{{Swiggum}
  et~al.}{2021}]{2021SwDO}
{Swiggum} C.,  et~al., 2021, arXiv e-prints, \href
  {https://ui.adsabs.harvard.edu/abs/2021arXiv210110380S} {p. arXiv:2101.10380}

\bibitem[\protect\citeauthoryear{{Sz{\"o}lgy{\'e}n}, {Meiron}  \&
  {Kocsis}}{{Sz{\"o}lgy{\'e}n} et~al.}{2019}]{AnisotropicMS}
{Sz{\"o}lgy{\'e}n} {\'A}.,  {Meiron} Y.,   {Kocsis} B.,  2019, \mn@doi [\apj]
  {10.3847/1538-4357/ab50bb}, \href
  {https://ui.adsabs.harvard.edu/abs/2019ApJ...887..123S} {887, 123}

\bibitem[\protect\citeauthoryear{Theiler}{Theiler}{1990}]{fractalknn}
Theiler J.,  1990, \mn@doi [J. Opt. Soc. Am. A] {10.1364/JOSAA.7.001055}, 7,
  1055

\bibitem[\protect\citeauthoryear{{Tiongco}, {Vesperini}  \& {Varri}}{{Tiongco}
  et~al.}{2016a}]{2016aTiVe}
{Tiongco} M.~A.,  {Vesperini} E.,   {Varri} A.~L.,  2016a, \mn@doi [\mnras]
  {10.1093/mnras/stv2574}, \href
  {https://ui.adsabs.harvard.edu/abs/2016MNRAS.455.3693T} {455, 3693}

\bibitem[\protect\citeauthoryear{{Tiongco}, {Vesperini}  \& {Varri}}{{Tiongco}
  et~al.}{2016b}]{frac_ret}
{Tiongco} M.~A.,  {Vesperini} E.,   {Varri} A.~L.,  2016b, \mn@doi [\mnras]
  {10.1093/mnras/stw1341}, \href
  {https://ui.adsabs.harvard.edu/abs/2016MNRAS.461..402T} {461, 402}

\bibitem[\protect\citeauthoryear{{Tiongco}, {Vesperini}  \& {Varri}}{{Tiongco}
  et~al.}{2017}]{2017TiVe}
{Tiongco} M.~A.,  {Vesperini} E.,   {Varri} A.~L.,  2017, \mn@doi [\mnras]
  {10.1093/mnras/stx853}, \href
  {https://ui.adsabs.harvard.edu/abs/2017MNRAS.469..683T} {469, 683}

\bibitem[\protect\citeauthoryear{{Tiongco}, {Vesperini}  \& {Varri}}{{Tiongco}
  et~al.}{2019}]{2019TiVe}
{Tiongco} M.~A.,  {Vesperini} E.,   {Varri} A.~L.,  2019, \mn@doi [\mnras]
  {10.1093/mnras/stz1595}, \href
  {https://ui.adsabs.harvard.edu/abs/2019MNRAS.487.5535T} {487, 5535}

\bibitem[\protect\citeauthoryear{{Trenti} \& {van der Marel}}{{Trenti} \& {van
  der Marel}}{2013}]{NoEnEq}
{Trenti} M.,  {van der Marel} R.,  2013, \mn@doi [\mnras]
  {10.1093/mnras/stt1521}, \href
  {https://ui.adsabs.harvard.edu/abs/2013MNRAS.435.3272T} {435, 3272}

\bibitem[\protect\citeauthoryear{{Vasiliev}}{{Vasiliev}}{2019}]{2019Va}
{Vasiliev} E.,  2019, \mn@doi [\mnras] {10.1093/mnras/stz2100}, \href
  {https://ui.adsabs.harvard.edu/abs/2019MNRAS.489..623V} {489, 623}

\bibitem[\protect\citeauthoryear{{Vesperini}, {Varri}, {McMillan}  \&
  {Zepf}}{{Vesperini} et~al.}{2014}]{2014VeVa}
{Vesperini} E.,  {Varri} A.~L.,  {McMillan} S.~L.~W.,   {Zepf} S.~E.,  2014,
  \mn@doi [\mnras] {10.1093/mnrasl/slu088}, \href
  {https://ui.adsabs.harvard.edu/abs/2014MNRAS.443L..79V} {443, L79}

\bibitem[\protect\citeauthoryear{{Wan} et~al.,}{{Wan} et~al.}{2021}]{2021WaOl}
{Wan} Z.,  et~al., 2021, \mn@doi [\mnras] {10.1093/mnras/stab306}, \href
  {https://ui.adsabs.harvard.edu/abs/2021MNRAS.502.4513W} {502, 4513}

\bibitem[\protect\citeauthoryear{{Watkins}, {van der Marel}, {Bellini}  \&
  {Anderson}}{{Watkins} et~al.}{2015}]{2015Wava}
{Watkins} L.~L.,  {van der Marel} R.~P.,  {Bellini} A.,   {Anderson} J.,  2015,
  \mn@doi [\apj] {10.1088/0004-637X/803/1/29}, \href
  {https://ui.adsabs.harvard.edu/abs/2015ApJ...803...29W} {803, 29}

\bibitem[\protect\citeauthoryear{{Webb} \& {Vesperini}}{{Webb} \&
  {Vesperini}}{2017}]{eneq_webb}
{Webb} J.~J.,  {Vesperini} E.,  2017, \mn@doi [\mnras] {10.1093/mnras/stw2513},
  \href {https://ui.adsabs.harvard.edu/abs/2017MNRAS.464.1977W} {464, 1977}

\makeatother
\end{thebibliography}
\appendix
\section{Additional Simulations}
\label{Appendix}
In order to explore the extent of stochastic variations in the results presented in this paper, we have run two additional simulations  for different random realizations of the initial conditions of the \URHigh\ and \CRHigh\ models.
The following figures present the results concerning the main structural and kinematic properties discussed in the paper for these two simulations (hereafter we refer to these additional simulations as \URHighA\ and \CRHighA).
These figures illustrate the degree of variance associated with these simulations and confirm the presence of all the main dynamical features identified in the paper.

\setcounter{figure}{0}
\begin{figure*}
    \centering
    \includegraphics[width=0.4\textwidth]{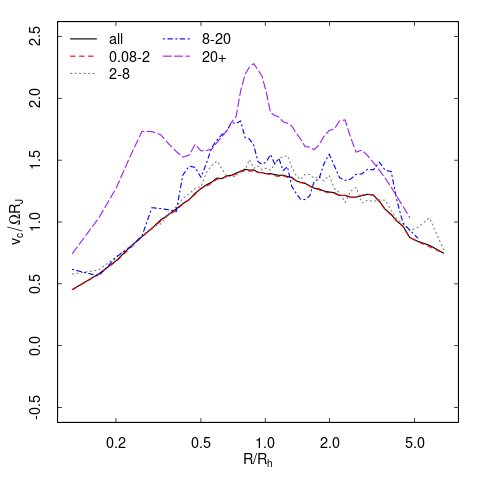}
    \includegraphics[width=0.4\textwidth]{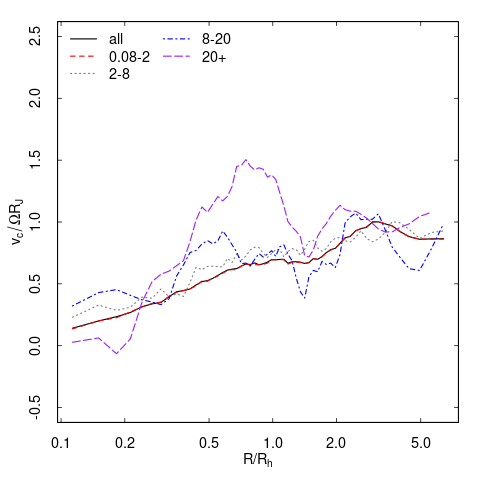}
    \caption{Same as middle row of Fig. \ref{fig:Vcurve_low} but for the simulations  \URHighA\ and \CRHighA.}
\end{figure*}

\begin{figure*}
    \centering
    \includegraphics[width=0.4\textwidth]{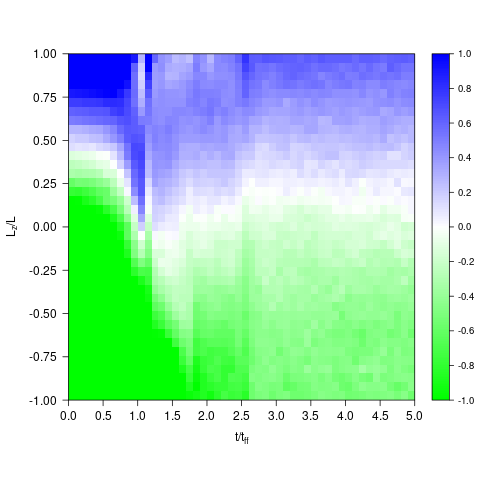}
    \includegraphics[width=0.4\textwidth]{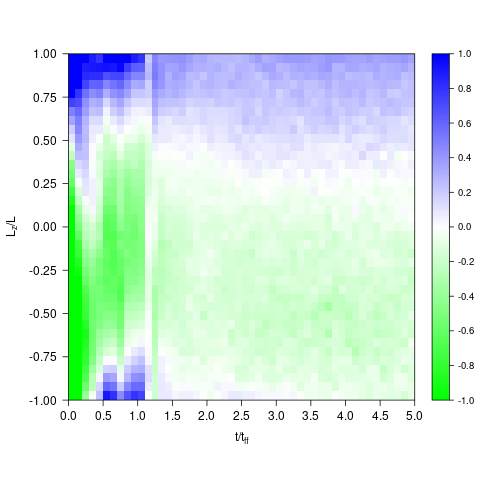}
    \caption{Same as Fig. \ref{fig:LzLEvo_all} but for the simulations \URHighA\ and \CRHighA}
\end{figure*}

\begin{figure*}
    \centering
    \includegraphics[width=0.4\textwidth]{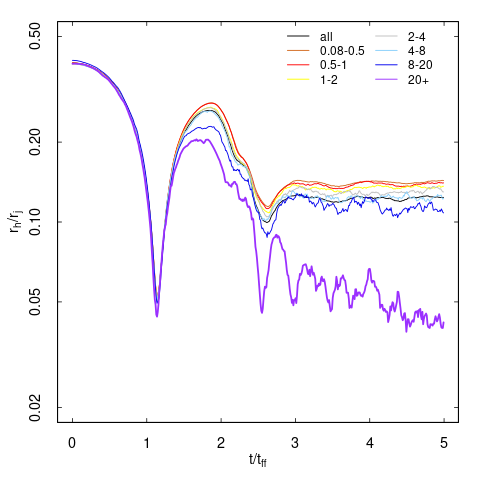}
    \includegraphics[width=0.4\textwidth]{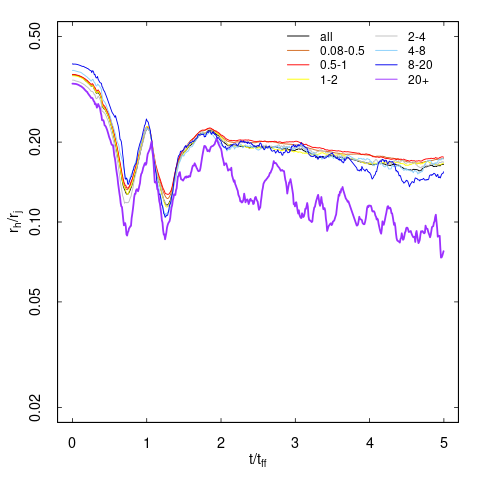}
    \caption{Same as Fig. \ref{fig:HMRadii} but for the simulations \URHighA\ and \CRHighA.}
\end{figure*}

\begin{figure*}
    \centering
    \includegraphics[width=0.4\textwidth]{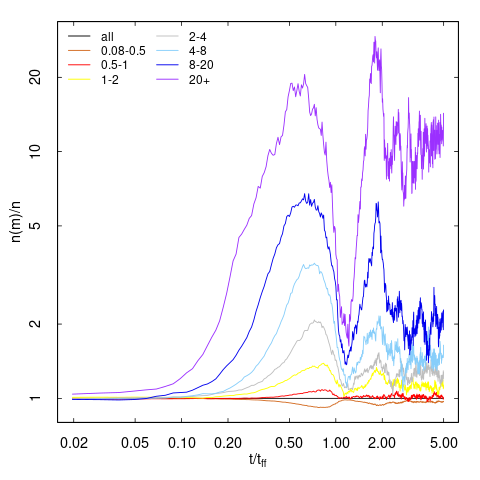}
    \includegraphics[width=0.4\textwidth]{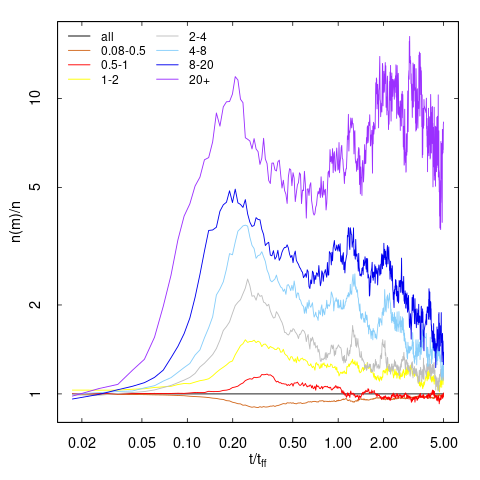}
    \caption{Same as Fig. \ref{fig:MSeg} but for the simulations \URHighA\ and \CRHighA.}
\end{figure*}

\begin{figure*}
    \centering
    \includegraphics[width=0.4\textwidth]{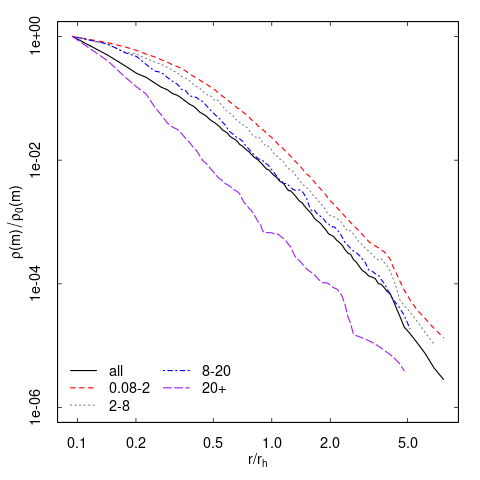}
    \includegraphics[width=0.4\textwidth]{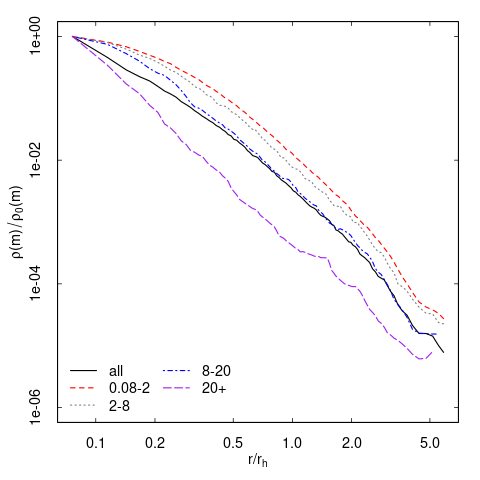}
    \caption{Same as Fig. \ref{fig:Dens_mass} but for the simulations \URHighA\ and \CRHighA.}
\end{figure*}

\begin{figure*}
    \centering
    \includegraphics[width=0.4\textwidth]{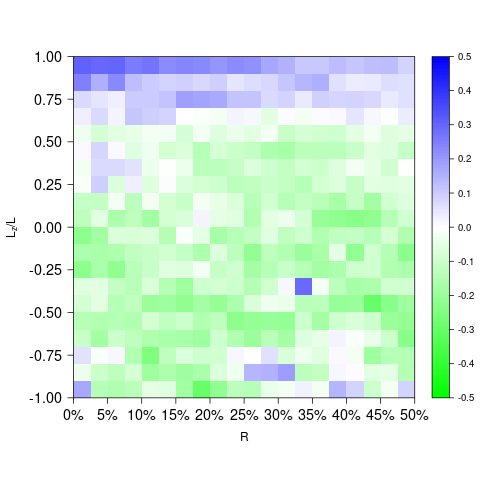}
    \includegraphics[width=0.4\textwidth]{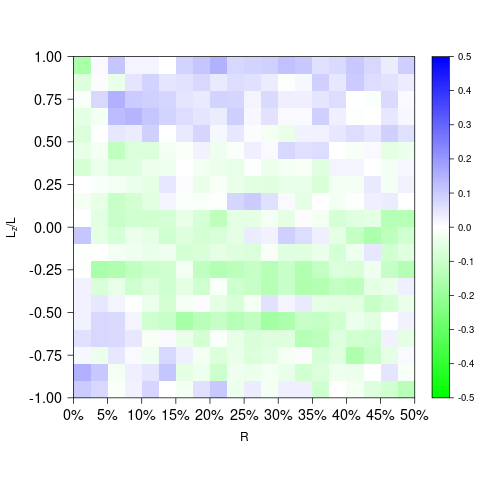}
    \caption{Same as Fig. \ref{fig:AvgMassEn} but for the simulations \URHighA\ and \CRHighA.}
\end{figure*}
\begin{figure*}
    \centering
    \includegraphics[width=0.4\textwidth]{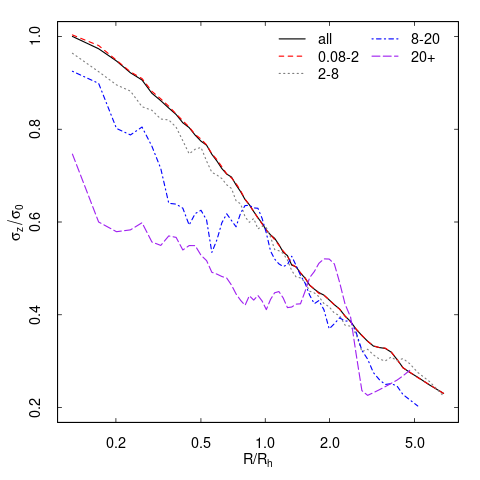}
    \includegraphics[width=0.4\textwidth]{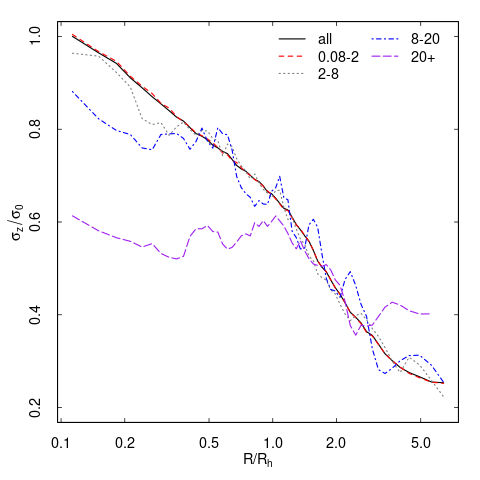}
    \caption{Same as Fig. \ref{fig:DispvsR} but for the simulations \URHighA\ and \CRHighA.}
\end{figure*}
\begin{figure*}
    \centering
    \includegraphics[width=0.4\textwidth]{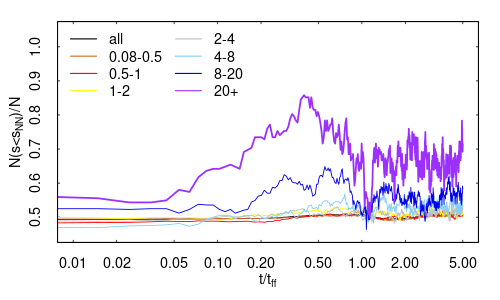}
    \includegraphics[width=0.4\textwidth]{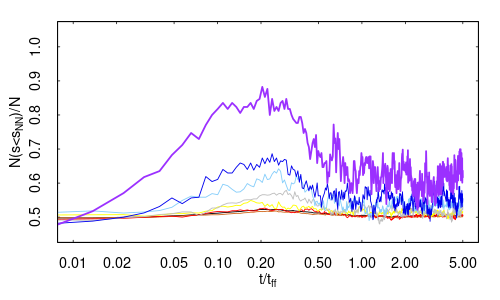}
    \includegraphics[width=0.4\textwidth]{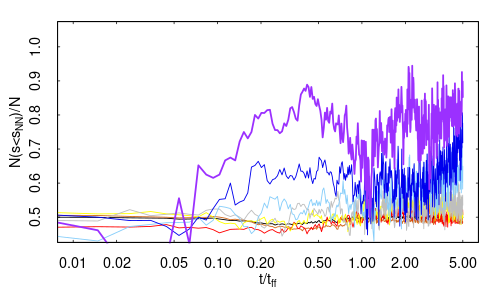}
    \includegraphics[width=0.4\textwidth]{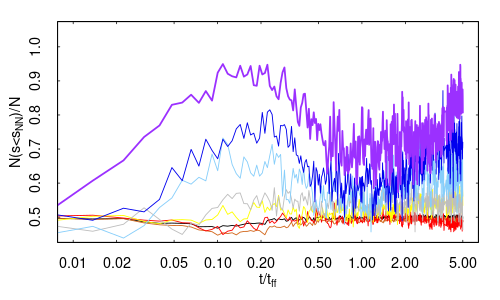}
    \caption{Same as Fig. \ref{fig:EEqui_localall} but for the simulations \URHighA\ and \CRHighA.}
\end{figure*}

\begin{figure*}
    \centering
    \includegraphics[width=0.4\textwidth]{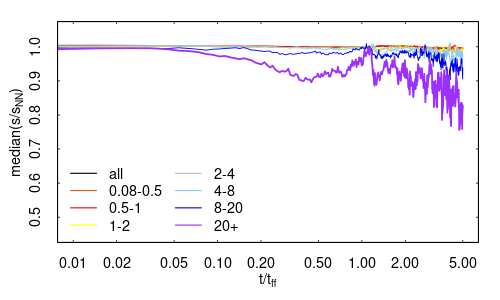}
    \includegraphics[width=0.4\textwidth]{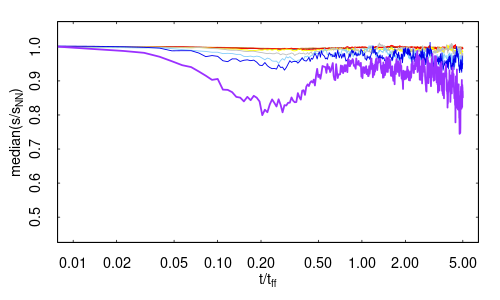}
    \includegraphics[width=0.4\textwidth]{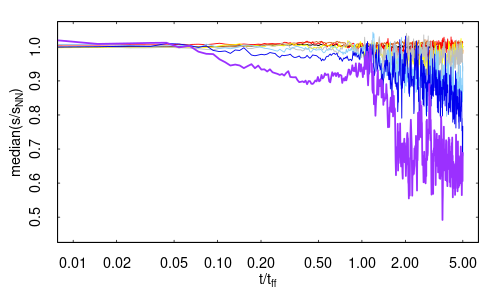}
    \includegraphics[width=0.4\textwidth]{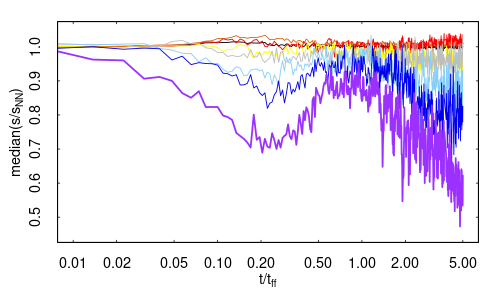}
    \caption{Same as Fig. \ref{fig:EEqui_div_localall} but for the simulations  \URHighA\ and \CRHighA.}
\end{figure*}

\bsp	
\label{lastpage}
\end{document}